\documentclass[graybox]{svmult}

\usepackage{type1cm}        
\usepackage{makeidx}         
\usepackage{graphicx}        
\usepackage{multicol}        
\usepackage[bottom]{footmisc}
\usepackage[scientific-notation=true]{siunitx}

\usepackage{newtxtext}       %
\usepackage[varvw]{newtxmath}       
\usepackage{xspace}
\usepackage{subfigure}
\usepackage{datatool}

\def\gwh{gravitational-wave\xspace}
\def\cwh{continuous-wave\xspace}
\def\tcwh{transient continuous-wave\xspace}
\def\gws{gravitational waves\xspace}

\def\nss{neutron stars\xspace}

\def\dm{dark matter\xspace}
\def\dmh{dark-matter\xspace}
\def\bh{black hole\xspace}
\def\bhs{black holes\xspace}
\def\ssm{sub-solar mass\xspace}
\def\pbh{primordial black hole\xspace}
\def\pbhs{primordial black holes\xspace}

\def\emri{EMRI\xspace}
\def\emris{EMRIs\xspace}
\def\LVK{LIGO, Virgo and KAGRA\xspace}
\def\fpbh{f_{\rm PBH}}
\def\mf{matched filtering\xspace}

\def\fmin{f_{\rm min}}
\def\fmax{f_{\rm max}}
\def\Tobs{T_{\rm obs}}
\def\Tfft{T_{\rm FFT}}
\def\l{\left(}
\def\r{\right)}

\def\snr{signal-to-noise ratio\xspace}
\newcommand{\msun}{\ensuremath{M_\odot}\xspace}

\makeindex             

\DTLsetseparator{,}

\newcommand{\avgVT}{\ensuremath{\left\langle VT \right\rangle}}

\newcommand{\bea}{\begin{eqnarray}}
\newcommand{\eea}{\end{eqnarray}}
\newcommand{\be}{\begin{equation}}
\newcommand{\ee}{\end{equation}}

\begin{document}

\title*{Gravitational waves from sub-solar mass Primordial Black Holes
}
\author{Andrew L. Miller}
\institute{Andrew L. Miller \at Nikhef -- National Institute for Subatomic Physics,
Science Park 105, 1098 XG Amsterdam, The Netherlands \\ and \\ Institute for Gravitational and Subatomic Physics (GRASP),
Utrecht University, Princetonplein 1, 3584 CC Utrecht, The Netherlands \\ \email{andrew.miller@nikhef.nl}}
%
%
\maketitle

\abstract{Gravitational waves from inspiraling sub-solar mass compact objects would provide almost definitive evidence for the existence of primordial black holes. In this chapter, we explain why these exotic objects are interesting candidates for current and future gravitational-wave observatories, and provide detailed explanations of how they are searched for. We describe one method, matched filtering, to search for binaries with masses between $[0.01,1]M_\odot$. Furthermore, since signals from inspiraling planetary- and asteroid-mass mass compact binaries ($[10^{-9},10^{-2}]M_\odot$) would spend hours to years in the detector frequency band, we explain the novel pattern recognition techniques that have been developed to search for them. Finally, we describe extreme mass ratio inspiral (EMRI) systems, and how these will be searched for in future space-based detectors. For all mass regimes, we comment on the prospects for detection.  }

\section{Introduction}
\label{sec:subolar_intro}

Sub-solar mass compact objects are broadly defined as objects more compact than white dwarfs with masses less than a solar mass. Predictions as far back as the 1930s by Chandrasekhar dictate an upper limit of the mass of a white dwarf to be $M_{\rm max}\sim 1.4M_\odot$ \cite{Chandrasekhar:1931ih,Chandrasekhar:1935zz,Suwa:2018uni,Muller:2018utr,Ertl:2019zks}, thus  any system with a mass larger than $M_{\rm max}$ would inevitably collapse to a neutron star or \bh. Since there are not any astrophysical channels to form sub-solar mass objects more compact than white dwarfs -- though neutron stars with masses between $[0.1,1]M_\odot$ are indeed stable for certain equations of state and some evidence has been presented for the existence of such objects \cite{2022NatAs...6.1444D} --, a detection of a sub-solar mass object via \gws could imply a new, unknown formation channel, one of which could be through accumulated \dm resulting in collapse \cite{DAmico:2017lqj,Shandera:2018xkn,Choquette:2018lvq}.

Data from ground-based \gwh observatories has extensively been used to search looking for \gws from \nss and \bhs with at least one solar mass, which has resulted in $\mathcal{O}(100)$ of them \cite{KAGRA:2021vkt,Nitz:2021zwj,Olsen:2022pin}. Already, there have been detections of unexpected systems, e.g. in the so-called ``lower-mass mass gap'' between $[2,5]M_\odot$ (GW190814, GW190425 and GW230529) \cite{LIGOScientific:2020zkf,LIGOScientific:2020aai,LIGOScientific:2024elc} that are heavier than the most massive pulsar in our Galaxy ($\sim 2.14M_\odot)$\cite{NANOGrav:2019jur}, though of comparable mass to that of pulsar PSR J1748-2021B and to that of the remnant of the neutron-star merger GW170817 \cite{LIGOScientific:2017vwq}. In this mass gap, there had simply been no observations of compact objects with these masses, though some stellar evolution models could predict this gap or a continuous distribution of masses; hence, possible explanations, e.g. hierarchical mergers \cite{Silsbee:2016djf,Fragione:2019zhm,Antonini:2017ash} or objects of primordial origin \cite{Carr:1975qj}, were invoked. Therefore, there is not a clear explanation of this mass gap, whether there is in fact one, and thus the \pbh hypothesis cannot be ruled out.
Furthermore, even the first detection of \gws, GW150914, indicated the existence of \bhs with masses of at least $\sim25M_\odot$ \cite{LIGOScientific:2016aoc}, which confirmed that \bhs could both form in binaries and merge within a Hubble time.

Based on the unexpected physics we have already learned from ``canonical'' \gwh sources, the currently \emph{unknown} sources of \gws could open new windows into the history of the universe and the formation of (primordial) \bhs. 

The range of \ssm \pbhs is vast, spanning from $\sim [10^{-18},1]\msun$. If such objects form binary systems, they will emit \gws as they inspiral. The amount of time that the signals spend in-band also vary significantly as a function of the mass of the \pbhs -- lighter systems inspiral for much longer than heavier ones, since the duration scales inversely with the chirp mass of the binary -- thus, different methods have been designed to cover different mass regimes. In fact, interest in detecting sub-solar mass compact object began decades ago, with studies that showed in the advanced detector era, inspiraling MACHOs could be observed, or constraints could be placed on their existence \cite{Nakamura:1997sm,Ioka:1998nz}. The ideal signal processing technique, the matched filter \cite{Wainstein:1962vrq}, is tractable around $\mathcal{O}(0.1)\msun$ and above, but becomes too computationally heavy to be used below that. Therefore, other techniques that can handle longer-duration systems, i.e. those of masses $[10^{-5},10^{-2}]\msun$, lasting hours-days, have been developed to search for such systems. We call the signals emitted by these binaries ``transient continuous waves'' (tCWs). Systems with \pbhs with masses between $[10^{-10},10^{-5}]\msun$ last for durations much longer than the observation time of the detector, and thus methods developed to handle quasi-monochromatic, quasi-infinite signals can be employed here. These signals are called ``continuous waves'' (CWs).

\begin{figure}
    \centering
    \hspace{-3mm} 
    \includegraphics[width=\textwidth]{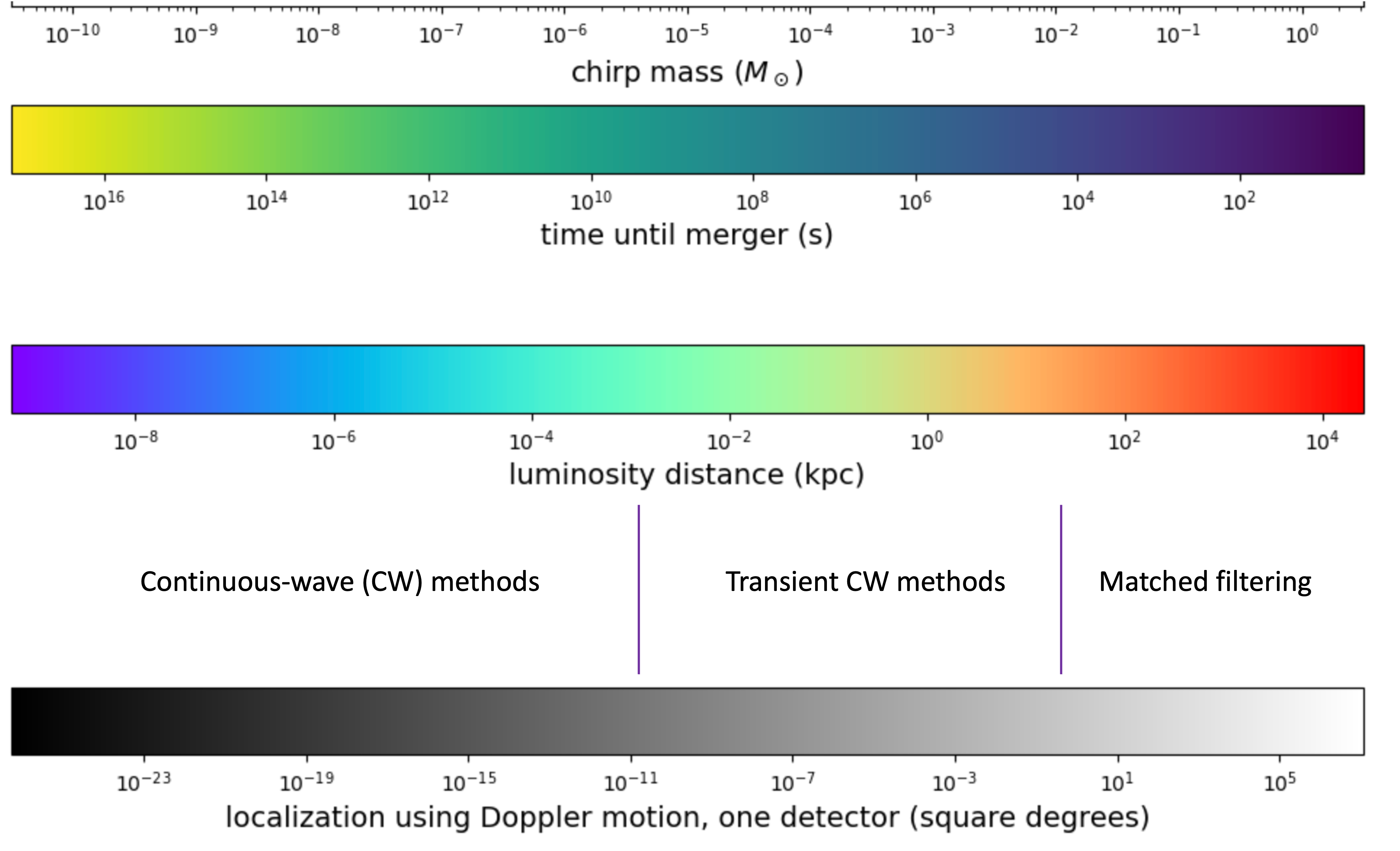}  
    \caption{Summary plot that shows the full sub-solar mass range that can be probed by ground-based \gwh detectors, along with different methods that can be used, the expected source durations, expected distance reaches, and maximal possible sky localization.}
    \label{fig:summ}
\end{figure}

In Fig. \ref{fig:summ}, we show the range of masses probed via \gws, the kinds of methods that can be used, the durations of the signals, the expected luminosity distance assuming strain sensitivities of $10^{-26},5\times 10^{-24},10^{-22}$ at 100 Hz in each of the three regimes (CW methods, tCW methods and matched filtering, respectively) and the maximal possible sky localization we could obtain using just by using the Doppler shift induced by the relative motion of the earth and the source at one detector. The values on each colorbar are meant to conceptually illustrate the different regimes in which we can search for \pbhs and should not be interpreted as strictly corresponding to the chirp mass listed, since, for example, luminosity distance reach depends on the actual strain sensitivity of the detector, and the time to merger is affected by the choice of frequency.

In this chapter, we will discuss the motivations for the existence of \ssm objects, the \gwh signatures of \ssm objects, and the techniques that we can use to detect them.

\section{Searches for GWs from $[0.1,1]M_\odot$ binary PBHs}
\label{sec:subsolar_0.1}

\subsection{Motivation}
\label{subsec:2}

Below a solar mass, black holes could be primordial in origin \cite{Carr:1975qj}. In essence, over-densities (relative to a background density) generated in the early universe, e.g. by large primordial fluctuations during inflation \cite{Carr:1993aq,Ivanov:1994pa,Kim:1996hr,Garcia-Bellido:1996mdl}, through bubble nucleation or domain walls \cite{Garriga:2015fdk}, or due to cosmic string loops and scalar field instabilities \cite{Khlopov:1985jw,Cotner:2017tir}, could collapse in on themselves, leading to the formation of ultra-compact objects \cite{Zeldovich:1967lct,Hawking:1971ei,Carr:1975qj,Chapline:1975ojl}. Because 84\% of the matter in the Universe is currently unexplained \cite{Planck:2018vyg}, \pbhs could comprise a fraction or all of this unknown matter. Depending on when they formed in the early universe, \pbhs could take on a wide range of masses -- as small as the Planck mass, $10^{-5}$ g if they formed at the Planck time ($10^{-43}$ s), $\sim 1M_\odot$ if they formed during the QCD phase transition ($10^{-5}$ s), or $10^5 M_\odot$ if they formed 1 second after the Big Bang \cite{Carr:1975qj}. The time and mass relations are obtained by equating the mean densities of a black hole of mass $M$ and radius $R$ and of the universe in the radiation era \cite{Villanueva-Domingo:2021spv}:

\begin{eqnarray}
    \rho_{\rm PBH} = \frac{M}{\frac{4}{3}\pi R_s^3} = \rho_{c} \sim 10^6 \left(\frac{t}{1\text{ s}}\right)^{-2} \text{ g/cm$^3$},
\end{eqnarray}
where $R_s = \frac{2 G M}{c^2} \sim 3 \left(\frac{M}{M_\odot}\right)$ km. Since the mass of the \pbh is roughly equivalent to the mass within the horizon, any fluctuations that enter the horizon cause over-densities and subsequent collapses into \pbhs.

Furthermore, the thermal of the universe could enhance \pbh formation at different scales \cite{Carr:2019kxo}, e.g. the QCD transition significantly reduces the radiation
pressure of the plasma, so that a uniform primordial enhancement stretching across the QCD scale will generate a distribution of \pbh masses that is sharply peaked around a solar
mass \cite{Byrnes:2018clq} as well as a broader mass distribution at both larger and \ssm masses \cite{LIGOScientific:2022hai}.

Other physical mechanisms could result in the formation of \ssm \pbhs. For example, if \dm is composed of particles that allow for dissipation or other chemical reactions, \ssm objects could form through collapsing \dmh halos \cite{DAmico:2017lqj,Shandera:2018xkn,Choquette:2018lvq}. Essentially, if a new ``dark Hydrogen atom'' exists, along with a heavy dark fermion (akin to a proton) with mass $m_x$, a light dark fermion (akin to an electron) with mass $m_c$, and a massless dark photon that mediates particle interactions in this theory with a strength determined by the dark fine structure constant $\alpha_D$, the dark Chandrasekhar mass upper limit becomes \cite{Chandrasekhar:1931ih,Shandera:2018xkn}:

\begin{equation}
    M_{\rm DC}=1.4M_\odot\left(\frac{m_p}{m_x}\right)^2 \label{eqn:Mdc}
\end{equation} 
and goes up to $\sim 1.9M_\odot$ for arbitrarily more complex \dmh particle interaction models \cite{Singh:2020wiq}. With the addition of this dark sector of particles, \dmh structures can undergo dissipative processes such as recombination, Bremsstrahlung radiation, and collisional excitation of atoms in sufficiently dense regions \cite{Buckley:2017ttd,Rosenberg:2017qia}, can clump like ordinary Hydrogen and therefore collapse into compact objects \cite{Shandera:2018xkn}. Without any dark nuclear physics, e.g. dark electron or neutron degeneracy pressures, all clumps of \dm will collapse into \bhs, with a minimum mass given by Eq. \ref{eqn:Mdc}. If $m_x>m_p$, \ssm ultra-compact objects could be formed.

While extensive searches for compact objects above a solar mass have been performed by the \LVK collaborations \cite{2015CQGra..32g4001L,2015CQGra..32b4001A,Aso:2013eba} and others \cite{KAGRA:2021vkt,Olsen:2022pin}, only a small effort \cite{Magee:2018opb,Phukon:2021cus,Nitz:2021mzz,LIGOScientific:2021job,LIGOScientific:2022hai,Nitz:2021vqh,Nitz:2022ltl,Miller:2024fpo} in terms of method design has been devoted to explore the full range of \ssm objects. There are many reasons for this, that will be explored in detail in the next sub-sections: (1) the \gwh amplitude falls of steeply with the masses of the two compact objects \cite{maggiore2008gravitational}, (2) searches for \ssm objects between $[0.1,1]M_\odot$ are computationally expensive due to need to correlate many long-duration waveforms with the data \cite{Magee:2018opb,Nitz:2022ltl}, (3) searches for \ssm objects below $0.1M_\odot$ require techniques beyond the standard \mf ones to handle long-duration signals in the detector \cite{Miller:2020kmv,Andrés-Carcasona:2023df,Alestas:2024ubs}, and (4) other existing experiments put stringent constraints on $\fpbh$ \cite{Green:2020jor}.

\subsection{Matched filtering searches}

Matched filtering searches correlate strain data with a
bank of deterministic templates, i.e. ``waveforms'', that  model \gwh emission from two compact objects of mass $m_1$ and $m_2$ in a quasi-circular orbit in many cases \cite{LIGOScientific:2019hgc}, though there are exceptions \cite{Nitz:2021mzz,Nitz:2022ltl}. For each template, maximized over arrival time, phase and distance to the source, a signal-to-noise ratio $\rho$ is computed, given by \cite{Babak:2012zx}:

\begin{equation}
    \rho^2 = \frac{\langle d| h \rangle^2 + \langle d | h_{\pi/2} \rangle^2}{\langle h | h \rangle}
    \label{eqn:mf}
\end{equation}
where
\begin{equation}
     \langle a | b \rangle \equiv 4\text{ Re}\left[\int_{f_{\rm min}}^{f_{\rm max}} df \frac{\tilde{a}^*(f)\tilde{b}(f)}{S_n(f)}\right]
    \label{eqn:inner}
\end{equation}
is the noise-weighted inner product, the tilde denotes the Fourier Transform of either the strain time-series $d$ or the template time-series $h$, ``$*$'' denotes the complex conjugate, $S_n(f)$ is the frequency-dependent power spectral density of the noise, estimated empirically using, typically, a  median-average \cite{Capano:2017fia}, and $\fmin$ and $\fmax$ are the minimum and maximum frequencies of the analysis. In the stationary-phase approximation for waveforms in the frequency-domain, $\tilde{h}_{\pi/2}=i\tilde{h}$ \cite{Droz:1999qx}. If the data contain exactly the signal that we are searching for, i.e., $d=h$, we would obtain the optimal signal-to-noise ratio:

\begin{equation}
    \rho^2_{\rm opt} =4\int_{f_{\rm min}}^{f_{\rm max}} df \frac{|\tilde{h}(f)|^2}{S_n(f)}
    \label{eqn:mfopt}
\end{equation}

Because we do not know which \gwh signals are embedded in the data, we must convolve all possible waveforms with the data. In practice, since we do not have infinite computing power, we cannot actually do this, so we must construct a bank of templates across a high-dimensional parameter space in such a way as to ensure that we do not lose more than some percent of signals due to discretizing the parameter space.

Let us assume we have a template $u$ and we correlate it with noise $n$ plus signal $s$, $n+\mathcal{A}s$, where $\mathcal{A}$ quantifies the amplitude of the signal. The expectation value of matched-filter \snr is \cite{Owen:1995tm}:

\begin{eqnarray}
    E[\rho^2] &=& E[\langle n+\mathcal{A}s|u\rangle] = E[\langle n|u\rangle] + \mathcal{A} E[\langle s|u\rangle] \nonumber \\
    &=& \begin{cases}
    \mathcal{A},& \text{if } s=u\\
    \mathcal{A}\langle s|u\rangle,              & \text{otherwise},
\end{cases}
\end{eqnarray}
where $E[\langle n|u\rangle]=0$ because the noise and signal are uncorrelated.
Therefore, we can read off a quantity that defines the effectiveness of our template: the ``match'' $\langle s|u\rangle \in [0,1]$ between the signal $s$ in the data and the template $u$.

Not all parameters of $u$ carry equal importance in \gwh searches. There are two types of parameters to consider: the \emph{intrinsic}, i.e. phase, parameters $\vec{\lambda}$, of $u$, and the \emph{extrinsic} parameters $\vec{\mu}$, e.g. the observed time of coalescence or the amplitude. The key distinction is that the full range of extrinsic parameters can be searched over simultaneously for a fixed intrinsic parameter $\vec{\lambda}$; thus, what matters for the construction of template banks are the intrinsic parameters, i.e. the \emph{phase evolution} of a \gwh signal.

We are very sensitive to changes in phase evolution, so we can maximize over the \emph{extrinsic} parameters and place templates in the \emph{intrinsic} parameter space $\vec{\lambda}$. Assuming that the signal matches ones of the templates in our template bank, 
we can write the match between a template $u(\vec{\mu},\vec{\lambda})$ and $u(\vec{\mu}+\Delta\vec{\mu},\vec{\lambda}+\Delta\vec{\lambda}) $ as:

\begin{equation}
\label{eqn:match definition}
M(\vec{\lambda},\Delta\vec{\lambda})\equiv
\renewcommand{\arraystretch}{.6}
\begin{array}[t]{c}{\textstyle\max}\\{\scriptstyle\mu,\Delta\mu}\end{array}
\langle u(\vec{\mu},\vec{\lambda}) |
u(\vec{\mu}+\Delta\vec{\mu},\vec{\lambda}+\Delta\vec{\lambda}) \rangle .
\end{equation}
Noting that the best-case scenario is when $\Delta\vec{\lambda}=0$, we can Taylor expand about $\Delta\vec{\lambda}$ and write:

\begin{equation}
\label{eqn:match expansion}
M(\vec{\lambda},\Delta\vec{\lambda})\approx 1 +
\frac{1}{2}
\left(
\frac{\partial^2 M}{\partial\Delta\lambda^i \partial\Delta\lambda^j}
\right)
{\atop\scriptstyle\Delta\lambda^k = 0}
\Delta\lambda^i \Delta\lambda^j .
\end{equation}
We can see that a concept of ``mistmatch'' $1-M$ arises naturally that depends on a \emph{metric} $g_{ij}$, which quantifies how close two templates should be when allowing a certain mismatch 

\begin{equation}
\label{eqn:metric definition}
g_{ij}(\vec{\lambda}) = -\frac{1}{2}\left(
\frac{\partial^2 M}{\partial\Delta\lambda^i \partial\Delta\lambda^j}\right)
{\atop\scriptstyle\Delta\lambda^k = 0}
\end{equation}
So, we can see that the template mismatch equals the proper distance between them:
\begin{equation}
\label{eqn:mismatch}
1-M=g_{ij}\Delta\lambda^i\Delta\lambda^j .
\end{equation}
Therefore, by taking derivatives of $M$ with respect to the physical parameters of the waveform, e.g. the chirp mass, frequency, or spins, we can efficiently place templates in the template bank for a given mismatch.

Typically, this template bank is constructed across a high-dimensional parameter space with a chosen maximum mismatch between adjacent templates of a few percent; in other words, no more than a few percent of signals will be lost due to a discrete sampling of the parameter space \cite{Owen:1995tm,Owen:1998dk}.

If we return to Eq. \ref{eqn:mf}, while $\fmax$ is typically taken to be $\sim 1024$ Hz, a choice based primarily on the fact that signal power accumulated above 1024 Hz is negligible. $\fmin$ is chosen to have a reasonable computational cost. This is because the number of templates scales inversely with \pbh parameters (e.g. minimum total mass $M_{\rm min}$, $f$, $M$) in the simple case of quasi-circular orbits with a given $S_n(f)$ \cite{Owen:1995tm}:  

\begin{equation}
\label{eqn:adv cal N}
{\cal N} \simeq
8.4\times 10^6\left(\frac{1-MM}{0.03}\right)^{-1}
\left(\frac{M_{\min}}{0.2~M_\odot}\right)^{-2.7} \left(\frac{f_0}{70\text{ Hz}}\right)^{-2.5}
\end{equation}
where $MM$ is the ``minimal match''. Thus, we see that the number of templates increases as the frequency and mass decrease. Physically, this is because of the time that the binary spends in the frequency band of the detector.
The time $\tau$ for a binary to reach coalescence is related to the frequency and chirp mass $\mathcal{M}\equiv\frac{(m_1m_2)^{3/5}}{(m_1+m_2)^{1/5}}$ \cite{maggiore2008gravitational}:

\begin{equation}
    \tau \simeq 2.18 \left(\frac{1.21M_\odot}{\mathcal{M}}\right)^{5/3}\left(\frac{100\text{ Hz}}{f}\right)^{8/3} \text{ s},
    \label{eqn:t-to-coal}
\end{equation}
and hence lower-frequency signals last longer and phase mismatches therefore have more time to accumulate. In other words, $\rho^2$ is sensitive to smaller changes in the signal
parameters, and the spacing between templates in the same parameter space must shrink to follow each potential waveform's phase evolution \cite{Owen:1995tm}. Thus, more templates are required to populate the parameter space at lower frequencies than at higher ones, and so $\fmin$ has been taken to be 45 Hz recently \cite{LIGOScientific:2021job,LIGOScientific:2022hai}, but could be taken to be lower.


While no search has yielded any viable \pbh candidates, upper limits on the fraction of \dm that \ssm \pbhs could compose have been computed. To do this, analyses inject many simulated signals into the data at different distances away, randomized over spins and all other parameters, at particular chirp masses and determine the detection efficiency at a particular confidence level.
%

\section{Even lighter PBHs}
\label{sec:3}

\subsection{Motivation} %

Primordial black holes need not lie within the region $[0.1,1]M_\odot$, but could take on even lighter masses as well. Recently detected star and quasar microlensing events ~\cite{Niikura:2019kqi,Hawkins:2020zie,bhatiani2019confirmation} suggest \pbhs could take on masses
$ [10^{-6}, 10^{-5}] M_\odot$ and constitute around 1\% of \dm, which agrees with the unified scenarios for \pbhs discussed in \cite{Clesse:2017bsw} but exceeds that predicted from floating planets \cite{KenathArun:2019jei}. Even Planet 9 could be a $10^{-6}M_\odot$ \pbh that the solar system captured long ago \cite{Scholtz:2019csj}. Though these observations provide a basis for the existence of \pbhs, astrophysical uncertainties, e.g. those due to clustering of \pbhs, are an unfortunate part of them \cite{Carr:2019kxo,Garcia-Bellido:2017xvr,Calcino:2018mwh,Belotsky:2018wph,Trashorras:2020mwn,DeLuca:2020jug}, which motivates the need for complementary probes of this mass range.

\subsection{Signal model}

For compact objects in binary systems with chirp masses $\mathcal{M}<0.1M_\odot$, we typically assume that the spin-up of the evolution over time $\dot{f}_{\rm gw}$ follows a simple power-law evolution, which is derived from the quadrupole formula by equating the orbital energy loss due to the \gwh power:

\begin{equation}
    \dot{f}_{\rm gw}=\frac{96}{5}\pi^{8/3}\left(\frac{G\mathcal{M}}{c^3}\right)^{5/3} f_{\rm gw}^{11/3},
    \label{eqn:fdot-pl}
\end{equation}
Thus, we only consider the inspiral portion of the system's life (neglecting merger and ringdown, due to the fact that the merger frequency lies well outside of the sensitivity frequency band of our detectors) and neglect higher-order contributions to the inspiral, i.e. we only care about the Post-Newtonian 0 contribution. Higher-order Post-Newtonian terms add negligibly to the phase for such light systems \cite{VelcaniThesis}, and \pbhs are expected to have low spins \cite{Clesse:2016vqa}; thus, the approximations made here should be valid.

Integrating Eq. \ref{eqn:fdot-pl}, we obtain the frequency evolution over time:

\begin{equation}
f_{\rm gw}(t)=f_0\left[1-\frac{8}{3}kf_0^{8/3}(t-t_0)\right]^{-\frac{3}{8}}~,
\label{eqn:f-of-t}
\end{equation}
where 
${k}\equiv\frac{96}{5}\pi^{8/3}\left(\frac{G\mathcal{M}}{c^3}\right)^{5/3}$, and the signal will have an amplitude of \cite{maggiore2008gravitational}:

\begin{equation}
h_0(t)=\frac{4}{d}\left(\frac{G \mathcal{M}}{c^2}\right)^{5/3}\left(\frac{\pi f_{\rm gw}(t)}{c}\right)^{2/3},
\label{eqn:h0(t)}
\end{equation}
where $d$ is the distance to the source.

\begin{figure}
    \centering
    \hspace{-3mm} 
    \includegraphics[width=\textwidth]{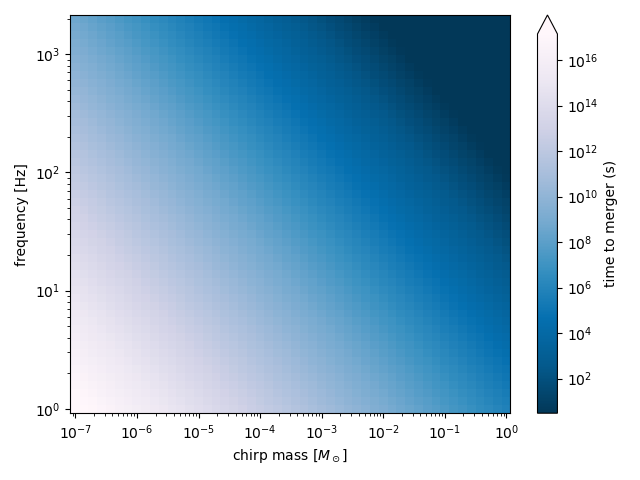}  
    \caption{Time before merger as a function of frequency and chirp mass. The widely distributed signal durations in the parameter space imply that different techniques are needed to probe the existence of PBHs at different masses. }
    \label{fig:tmerg}
\end{figure}

The time until coalescence, given by Eq. \ref{eqn:t-to-coal}, varies significantly across the mass and \gwh frequency parameter space, as can be seen in Fig. \ref{fig:tmerg}, based on \cite{Miller:2020kmv}. The long durations of these signals imply that \mf is impractical; thus, other methods had to be developed to probe such systems.

\begin{figure}
    \centering
    \includegraphics[width=\textwidth]{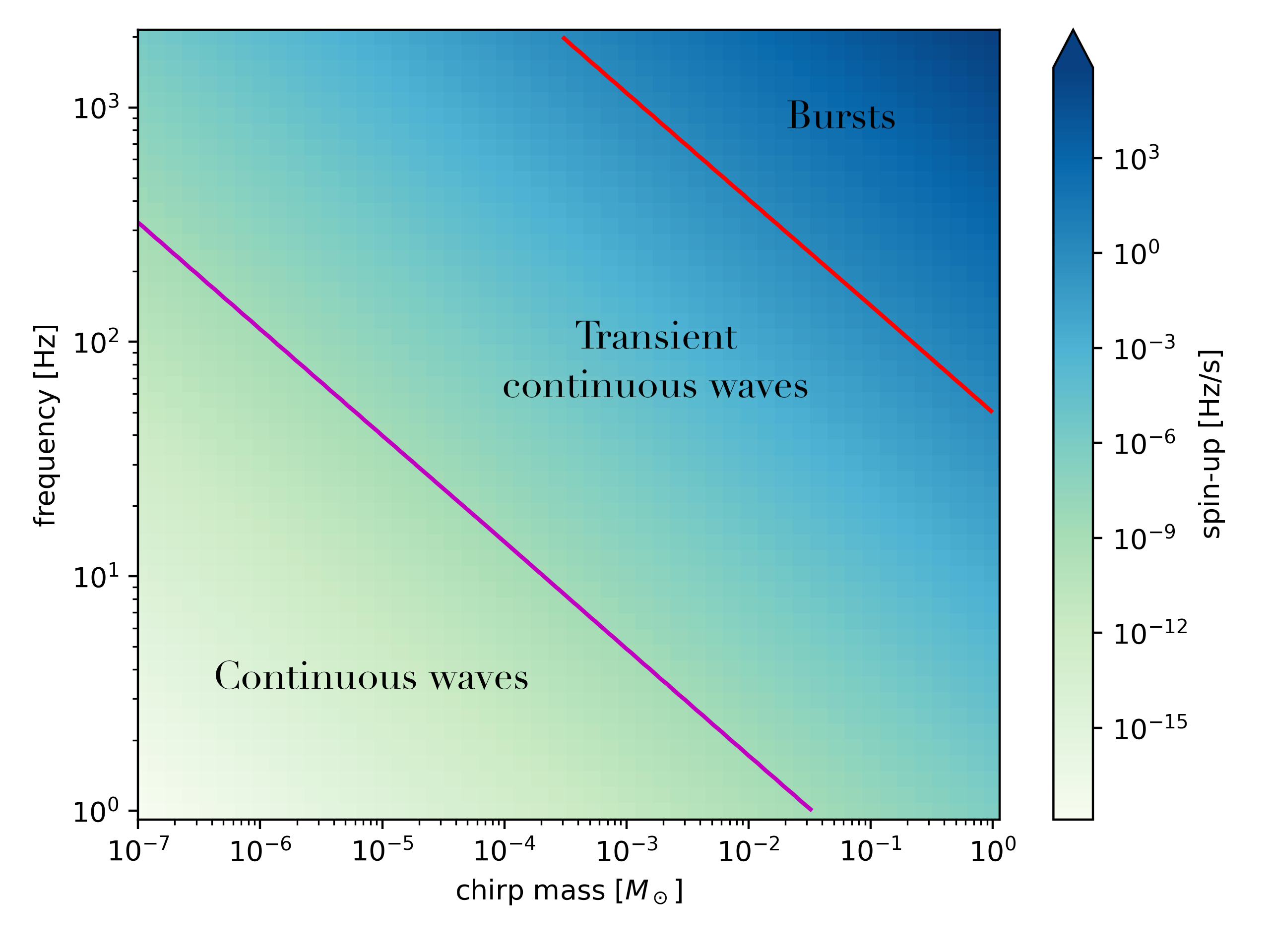}
    \caption{Frequency as a function of chirp mass with spin-up colored. A magenta line representing the maximum spin-up to which continuous-wave searches have considered is also plotted \cite{abbott2019all}, along with a red line indicating roughly the maximum spin-up that can be considered in a transient continuous-wave analysis (1 Hz/s). Colored points below the magenta line, meaning smaller spin-ups, represent possible masses of inspiraling PBHs that can be probed with continuous-wave methods. 
    Transient continuous gravitational-wave methods are necessary to exhaustively constrain larger PBH chirp masses.}
    \label{fig:fdotcw}
\end{figure}

We note that if the second term in Eq. \ref{eqn:f-of-t} is small compared to the first one, Eq. \ref{eqn:f-of-t} can be binomially expanded:

\begin{eqnarray}
    f_{\rm gw}(t)\simeq f_0\left[1+kf_0^{8/3}(t-t_0)\right] = f_0+\dot{f}(t-t_0),
    \label{eqn:cw}
\end{eqnarray}
which is nothing but a monochromatic \gwh signal whose frequency varies slowly in time by $\dot{f}$. Physically, such signals will have $\mathcal{M}<10^{-5}M_\odot$ and last for years, at least. This derivation suggests that we can apply methods to search for light \pbhs systems used in searches for ``continuous gravitational waves'', quasi-monochromatic, persistent signals canonically arising from lumpy, rotating neutron stars \cite{Tenorio:2021wmz,Riles:2022wwz,Piccinni:2022vsd,Miller:2023qyw}. When we must use the full power-law expression, we can, instead, apply methods used in searches for ``transient continuous waves'', canonically arising from newborn neutron stars with large deformations \cite{Riles:2017evm}. While the physics of these systems are all different, the frequency evolution over time follows the same equations. The divide between the \tcwh and the \cwh regimes is practically a function of the maximum $\dot{f}$ that \cwh searches probe, as well as the difference in frequency evolutions of Eq. \ref{eqn:f-of-t} and Eq. \ref{eqn:cw}, and can be seen in Fig. \ref{fig:fdotcw}. We see that \cwh searches are sensitive only to small chirp masses and slowly inspiraling systems, while \tcwh searches could be sensitive to a much larger parameter space that \mf cannot probe.

\subsection{Search techniques}\label{sec:meths}

Matched filtering techniques cannot be easily applied to search for these systems, since the computational cost of correlating numerous templates with the data scales strongly with the signal duration \cite{Miller:2023rnn}. Thus, techniques similar to those used to search for long-lived sources of continuous gravitational waves had to be developed to look for these systems. These methods are largely \emph{semi-coherent}, meaning that they break the dataset into chunks of length $\Tfft$, analyze the data within $\Tfft$ coherently (with phase information), and combine the data across different $\Tfft$ incoherently (only the power, no phase). These techniques are much more computationally efficient than matched filtering, and can be tuned to obtain similar sensitivities by setting lower thresholds on detection statistics to make up for the degradation in sensitivity induced by performing a semi-coherent analysis \cite{Prix:2009oha}. They also tend to be more robust against noise disturbances and are able to deal effortly with non-Gaussian, non-stationary noise \cite{Astone:2014esa,Riles:2022wwz}.

These techniques rest on the principle that, within each $\Tfft$, the frequency modulation of a \gwh signal is confined to one frequency bin $\Delta f = 1/\Tfft$. This means that within each $\Tfft$, the signal is monochromatic, though over $\Tobs$, the signal will be confined to different frequency bins. The question, then, is how to choose $\Tfft$. If $\dot{f}$ in Eq. \ref{eqn:fdot-pl} encapsulates the rate of change of the frequency, we would like to ensure that the following condition is met:

\begin{equation}
    \dot{f}\Tfft \leq \Delta f = \frac{1}{\Tfft}.
\end{equation}
Thus, in our case

\begin{equation}
    \Tfft \leq \sqrt{\frac{5}{96}}\pi^{-4/3}\left(\frac{G \mathcal{M}}{c^3}\right)^{-5/6} f_{\rm gw}^{-11/6}
    \label{eqn:tfftmax}
\end{equation}
Now, we know that $f_{\rm gw}$ may increase rapidly, which means that the maximum $\Tfft$ decreases. But, the sensitiivty of any given semi-coherent method depends on $\Tfft$. It is therefore worthwhile to consider ways to maximize the sensitivity as a function of $\Tobs$, $\Tfft$ and the noise power spectral density, which can be done analytically \cite{Alestas:2024ubs} or empirically \cite{Miller:2020vsl}.

Because inspiraling planetary-mass \pbhs last for durations in between \cwh sources and merging compact binaries, different techniques have to be used to be optimally sensitive to these systems. While methods have been developed to search for long-lived remnants of neutron star mergers or supernovae \cite{Sun:2018hmm,Oliver:2018dpt,Miller:2018rbg,Miller:2019jtp}, only one has been extensively adapted to search for these kinds of \pbhs \cite{Miller:2020vsl} that transforms points in the time/frequency plane of the detector to lines in the frequency/chirp mass plane of the source using the Hough transform \cite{Astone:2014esa,Miller:2018rbg},though others based on tracking time/frequency evolutions stochasticly using the Viterbi algorithm \cite{Viterbi:1967,Alestas:2024ubs} and demodulating the phase evolution \cite{Piccinni:2018akm,Andrés-Carcasona:2023df}, are being developed.

\subsubsection{Hough Transform}

\begin{figure*}[ht!]
     \begin{center}
        \subfigure[ ]{%
            \label{pm_injt}
            \includegraphics[width=0.5\textwidth]{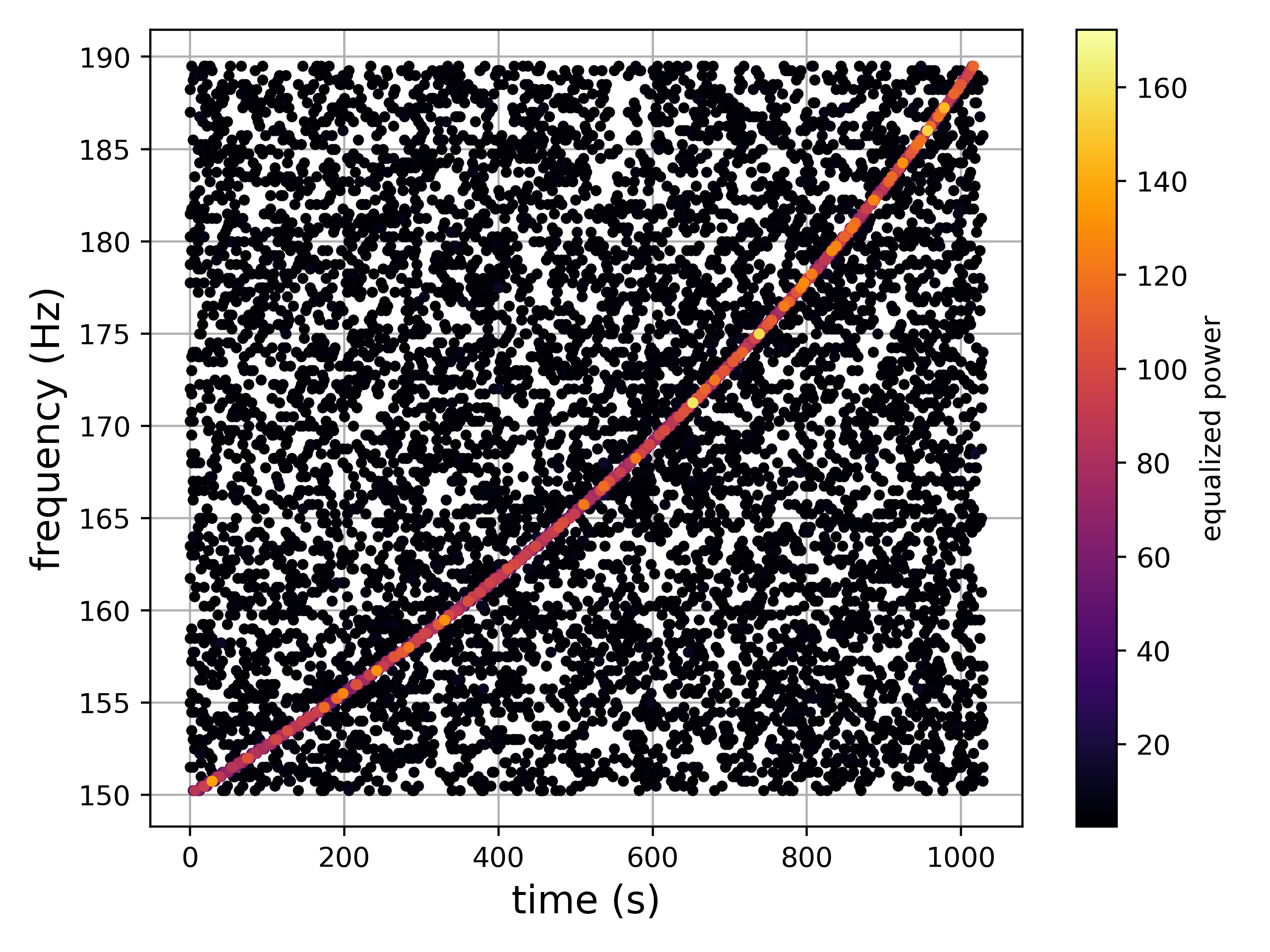}
        }%
        \subfigure[]{%
           \label{hm_inj}
           \includegraphics[width=0.5\textwidth]{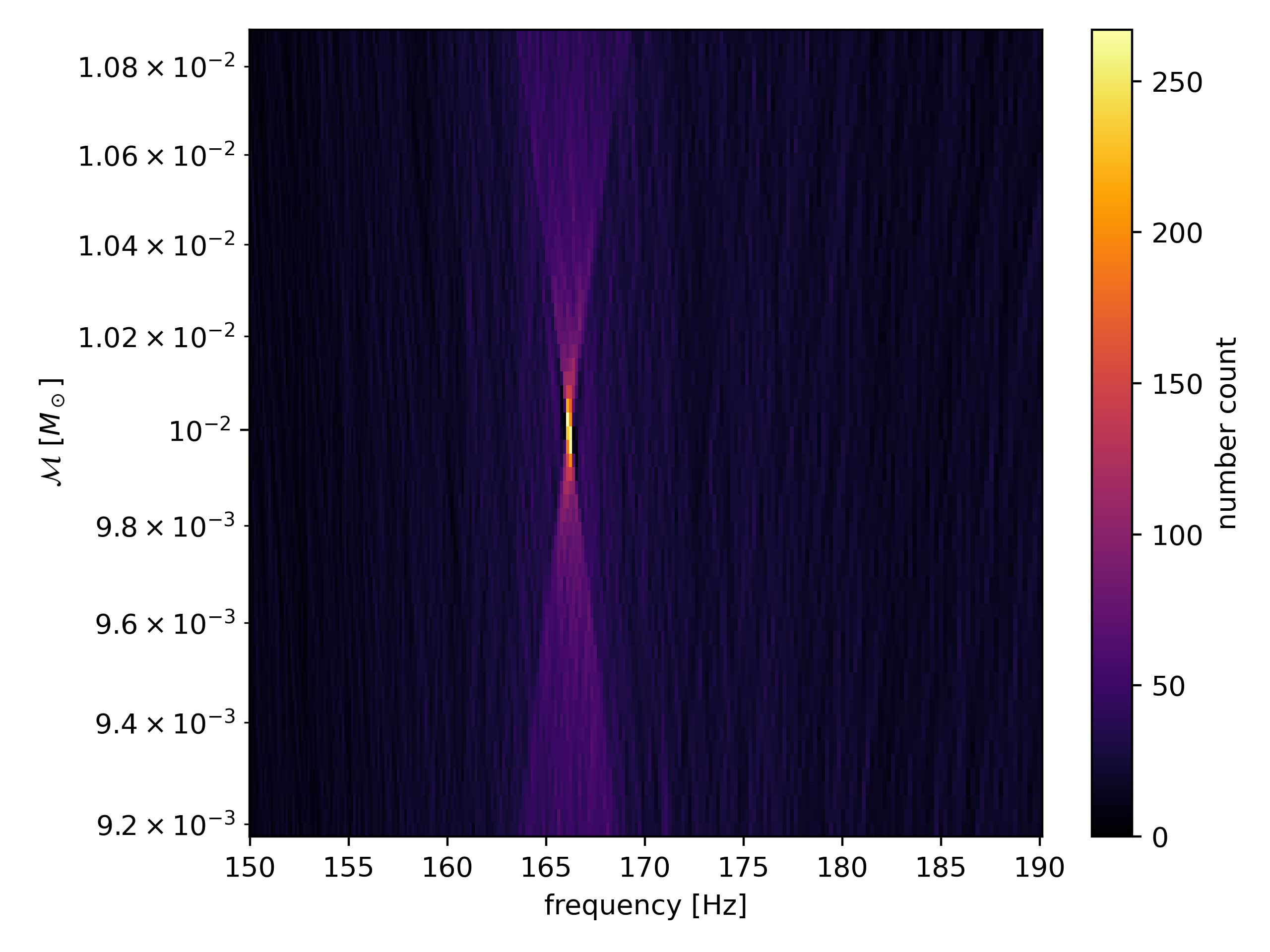}
        }\\ 
    \end{center}
    \caption[]{%
   The left-hand plot shows the peakmap (time/frequency map), created with $\Tfft=4$ s, of a strong, rapidly evolving signal, which is the input to the Generalized Frequency-Hough transform. The right-hand plot shows the output of the Generalized Frequency-Hough transform, which is a histogram in the $f_0$/$\mathcal{M}$ space of the source. The injection parameters are $h_0=10^{-22}$, $f_0\simeq165$ Hz, 
    $\mathcal{M}=10^{-2} M_\odot$. The recovered candidate is in the same bin as the injection.} 
   \label{fig:pm_hm_inj}
\end{figure*}

We show one example of a technique used to search for planetary-mass \pbhs in Fig. \ref{fig:pm_hm_inj}. Here, from the strain data $h(t)$, we take Fourier transforms of length $\Tfft$ over a duration of length $\Tobs$, estimate the power spectral density in each $\Tfft$, divide the square of the FFT by the estimate of the power spectral density, threshold the result, and create a time/frequency map, shown in the left panel of Fig. \ref{fig:pm_hm_inj}. This time/frequency map is the input to the Generalized frequency-Hough transform, which essentially loops over different chirp masses and sums over the times in the power law given in Eq. \ref{eqn:f-of-t}, and creates a two-dimensional histogram in a parameter space that relates to $f_0$ and $\mathcal{M}$, shown in the right panel of Fig. \ref{fig:pm_hm_inj}. This method does not actually sum the equalized power, but only a ``1'' when a pixel in the time/frequency plane is above a threshold, and zero otherwise. This is one example of how these semi-coherent methods deal with noise disturbances: in this case, powerful noise lines, regardless of their strength, are given weights of ``1'' at whatever times they appear, reducing their impact on the background.

\subsubsection{Viterbi}

Another method, used in \cwh all-sky searches for \gws from isolated neutron stars, can be applied to search for inspiraling \pbh binaries. This technique relies on the Viterbi
algorithm, which is a dynamical programming algorithm, to identify the most likely sequence of hidden Markov states given a set of observations \cite{Viterbi:1967}. Instead of computing the posterior probability that each track contains a signal and selecting that one the maximum, the method, at each time step in a set of observations, maximizes the probability to detect a \gwh signal. It essentially attempts to find the ``optimal path'' in a two-dimensional time/frequency representation of the data without looping through all the possible tracks \cite{Suvorova:2016rdc,Suvorova:2017dpm,Melatos:2021mmz}. In other words, tracks at one time step that will never become the one with maximum probability are rejected immediately.
Additionally, Viterbi treats the track as the result of a
Markovian process, i.e. the probability calculated at a given time/frequency point (a ``state'') depends only on the probability computed at the previous state. 

\begin{figure*}[t!]
\centering
\includegraphics[width=0.495\textwidth]{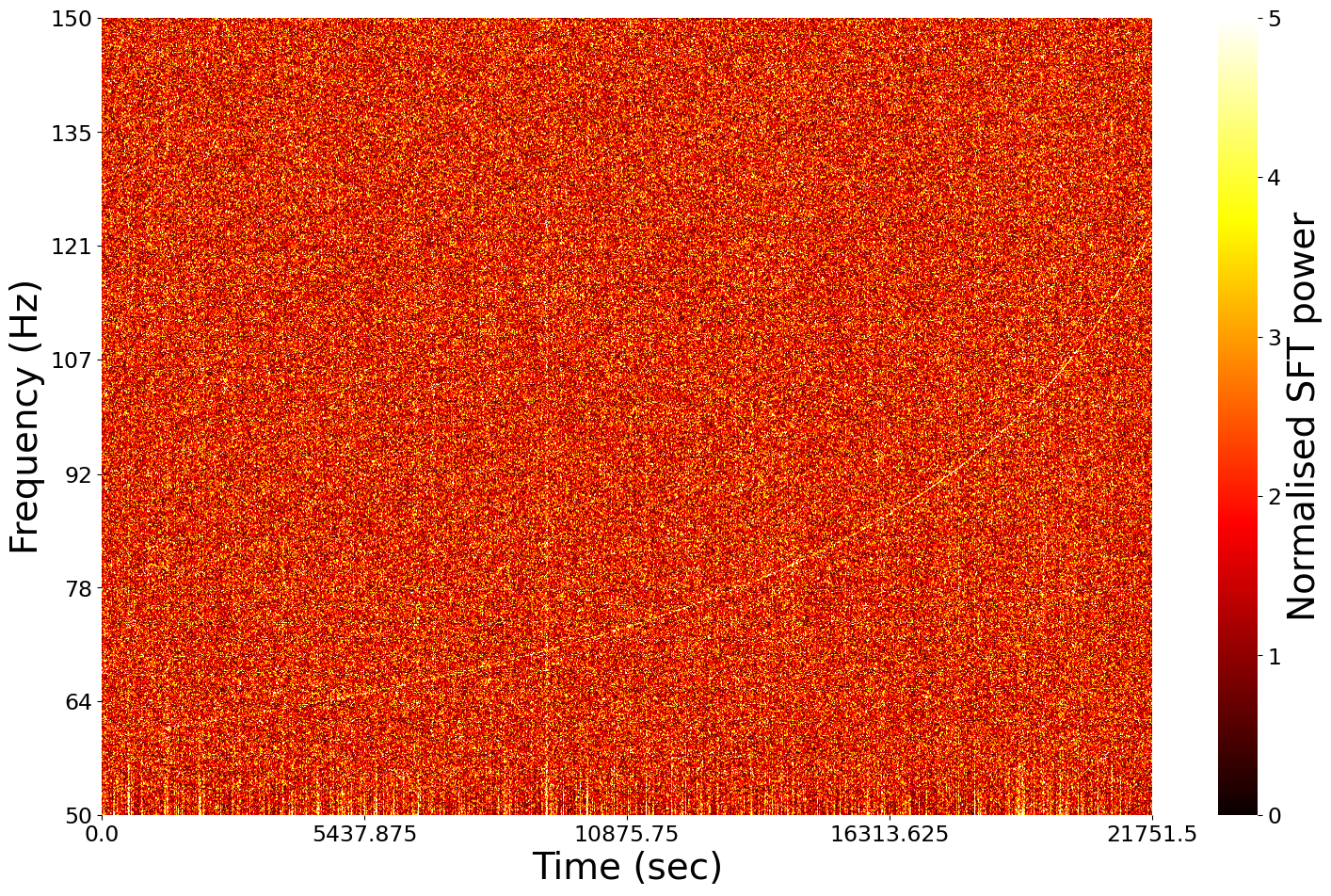}
\includegraphics[width=0.495\textwidth]{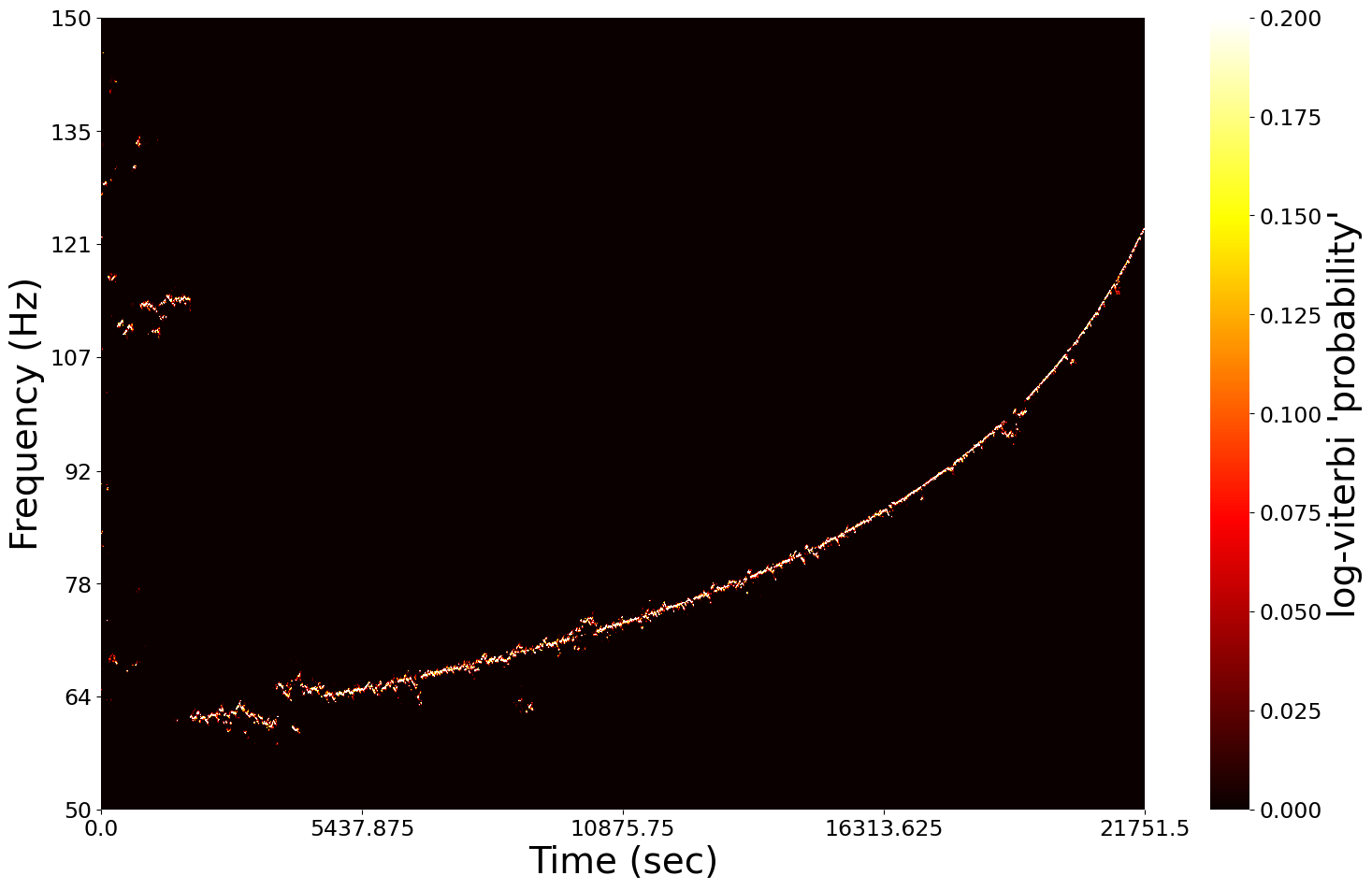}
\caption{\label{fig:signal_spec_det}Signal detection using the Viterbi algorithm in the case of $[\mathcal{M}_c, d_L] = [10^{-2} M_{\odot}, 147 \mathrm{kpc}]$. Left panel: The spectrogram of the injected signal hidden within the Gaussian noise. Right panel: The detected signal by the Viterbi algorithm. Taken from \cite{Alestas:2024ubs}. 
}
\end{figure*}

The Viterbi method is computationally cheap compared to the Hough Transform, since it operates in a model-agnostic way\footnote{In the case of inspiraling \pbhs, it is imposed in the prior probability that the signal frequency must increase over time, which is the only trace of ``modelling'' present in the method.} and does not require that each possible track is computed beforehand \cite{Bayley:2019bcb,Bayley:2020zfa,Bayley:2022hkz}. The trade-off, though, is a loss of sensitivity compared to the Hough. However, Viterbi would be sensitive to \pbhs for which the \gwh waveform deviates significantly from Eq. \ref{eqn:fdot-pl}, i.e. at higher Post-Newtonian orders required for highly asymmetric mass ratio systems \cite{Alestas:2024ubs}, or additional physics, e.g. \dmh clouds present around the \pbhs \cite{Brito:2015oca,Kavanagh:2020cfn,Coogan:2021uqv,Cole:2022ucw}. In Fig. \ref{fig:signal_spec_det}, we show how the Viterbi algorithm identifies and recovers the signal track present in the spectrogram.

\subsubsection{BSD-COBI}

The Band-Sampled Data COmpact Binary Inspiral (BSD-COBI) has also been designed to search for binary \pbhs \cite{Andrés-Carcasona:2023df}. This method relies on a concept known as ``heterodyning'', in which the \gwh frequency evolution over time is demodulated out of the data. The procedure essentially applies a phase correction to the strain time series\footnote{It's not the real-valued strain time series, but the so-called ``reduced analytic signal'' on which this correction is applied.} to remove a signal. 

In other words, consider that the strain $h(t)$ has the form

\begin{equation}
    h(t) =\text{Re}[A(t)e^{i\phi} + n(t)]
\end{equation}
where $n(t)$ is the noise, $A(t)$ is the time-varying signal amplitude, and $\phi$ is the phase of the signal (the integral of Eq. \ref{eqn:f-of-t}). At this point, we can multiply by a phase factor $e^{-\phi}$ to heterodyne or demodulate the data

\begin{eqnarray}
    h^{\rm het}(t) &=& \text{Re}[h(t)e^{-i\phi}] \\
    &=& \text{Re}[A(t)+n(t)]. \nonumber
\end{eqnarray}

If the correction is done perfectly, the signal does not experience any frequency modulation over $t$; thus, it becomes monochromatic over its total duration. In principle, the $\Tfft$ length can be increased to, at most, $\Tobs$; the sensitivity after this correction, therefore, rivals that of the matched filter. However, the signal parameters have to be exactly known to achieve matched filtering sensitivity. In practice, $\Tfft$ can be increased by a factor of $\mathcal{O}(1-10)$ relative to that in  Eq. \ref{eqn:tfftmax}, which enhances the sensitivity of the search. An efficient way of performing this correction, however, is under evaluation.

\begin{figure}
    \centering
    \includegraphics[width=\columnwidth]{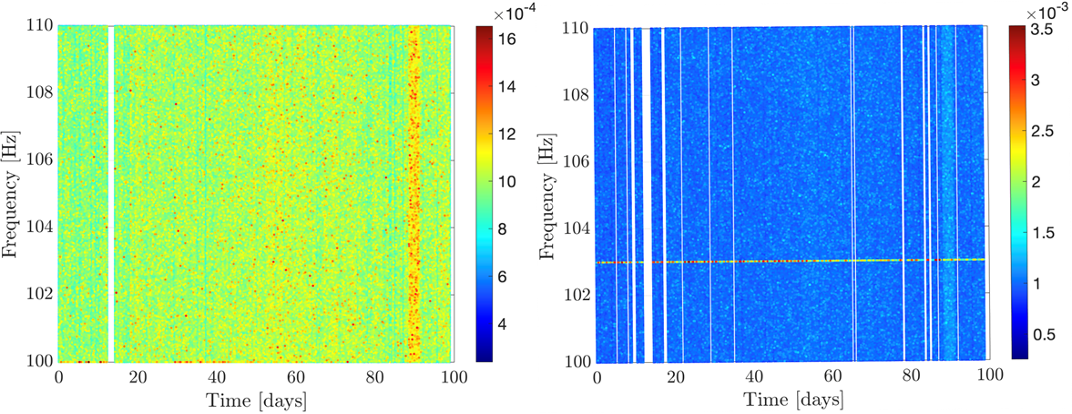}
    \caption{Left: original time/frequency peakmap, before heterodyning. The signal cannot be seen. Right: time/frequency map after heterodyning. All phase modulations have been removed, so the signal appears as monochromatic. Taken from \cite{Andrés-Carcasona:2023df}.}
    \label{fig:heterodyne}
\end{figure}

In Fig. \ref{fig:heterodyne}, we show the heterodyning procedure as applied to a \gwh signal from an inspiraling \pbh binary. We see that, before the correction, the signal cannot be seen in the time/frequency map, but after heterodyning, it becomes visible. At this point, the time/frequency map will be projected onto the frequency axis, which amounts to a binary sum over time of whether the power in each time/frequency pixel is above a given threshold.

\subsection{Constraints on asteroid mass \pbhs from all-sky searches for continuous waves}\label{sec:asteroid-mass-constraints}

While planetary-mass \pbhs have to be searched for with dedicated techniques, we can leverage existing all-sky searches for isolated, asymmetrically rotating neutron stars to place constraints on asteroid-mass \pbhs. This portion of the parameter space corresponds to the points below the light green line in Fig. \ref{fig:fdotcw}. Essentially, these searches \cite{KAGRA:2022dwb} look for spinning down (but also spinning up) isolated neutron stars, whose spin-down or spin-up can be explained by having mountains or accretion, respectively. Since \pbhs will inspiral, they will always be spinning up, hence it is the search over positive frequency drifts, up to $\dot{f}_{\rm max}\sim 2\times 10^{-9}$ Hz/s in \cite{KAGRA:2022dwb}, that allow us to make some statement on very slowly inspiraling \pbhs. The frequency range that was searched, from $\sim$[10,2000] Hz, the maximum spin-up, and the criteria that the frequency evolution be linear, $f(t)=f_0+\dot{f}(t-t_0)$, fix the chirp masses to which we would be sensitive based on Eq. \ref{eqn:fdot-pl} to be $\sim [10^{-7},10^{-5}]M_\odot$. 

The output of \cwh searches is often upper limits on the strain amplitude $h_0$ as a function of frequency, averaged over extrinsic parameters. From $h_0(f)$, we can compute the distance reach via Eq. \ref{eqn:h0(t)}, employing the upper limits obtained in \cite{KAGRA:2021una,KAGRA:2022dwb,Steltner:2023cfk}. 

\begin{figure}
    \centering
    \includegraphics[width=\columnwidth]{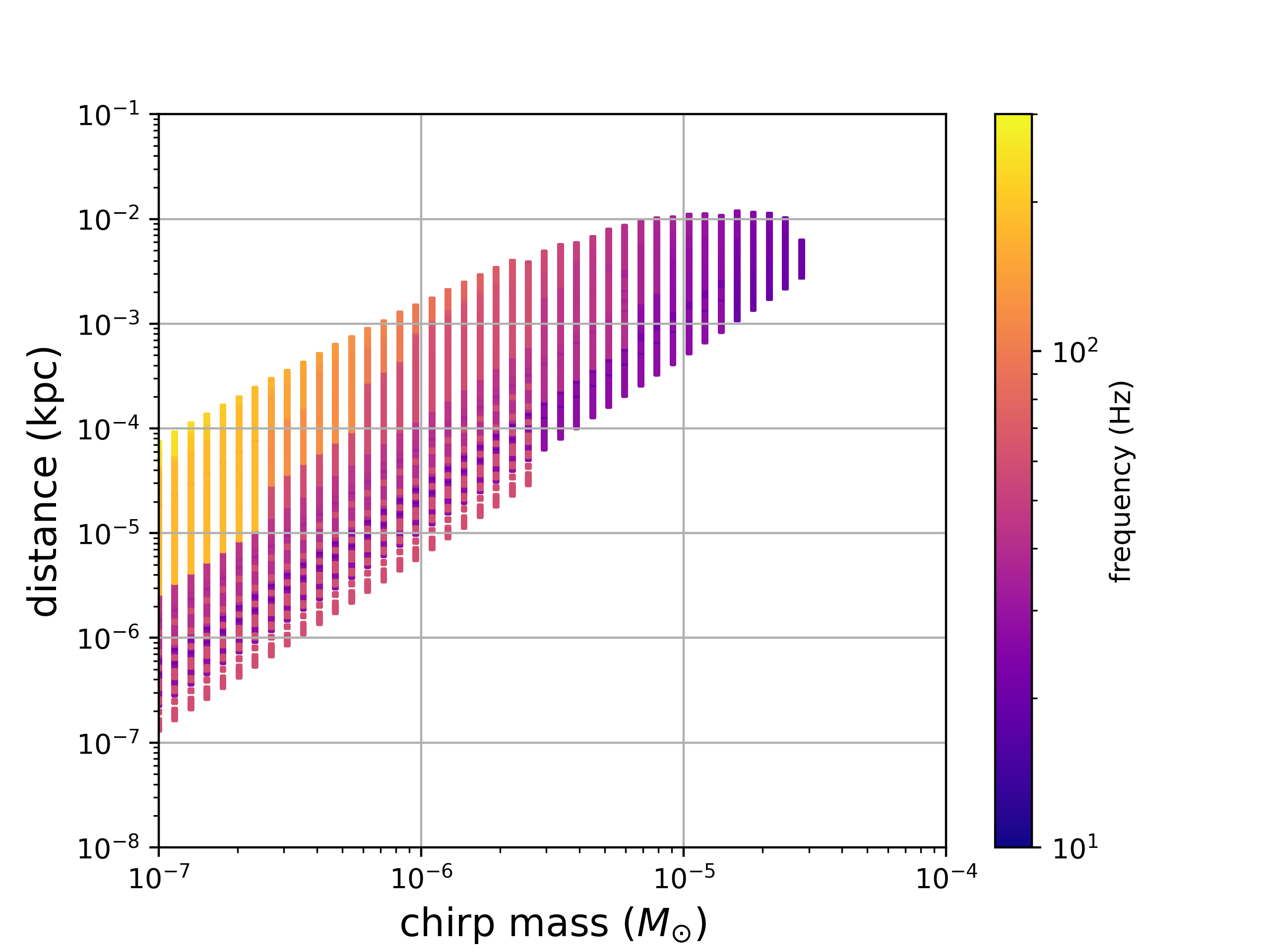}
    \caption{Distance reach as a function of chirp mass with frequency on the color axis. The maximum reach is of $\mathcal{O}(\rm pc)$ for most of the planetary-chirp-mass black hole inspirals. Each point contributes to equations \ref{eqn:nbin} and \ref{eqn:ntot} used to obtain constraints on the rates at a given chirp mass. Taken from \cite{Miller:2021knj}.}
    \label{fig:fhdist_constraints}
\end{figure}

We will now approximate theoretically the results we obtained in \cite{Miller:2021knj} for the distance reach as a function of chirp mass (Fig. \ref{fig:fhdist_constraints}), and the constraints on the fraction of \dmh that \pbhs could compose as a function of mass for equal-mass binaries, and as a function of $m_2$ for $m_1=2.5\msun$.

In \cwh semi-coherent searches, the upper limits on the strain amplitude as a function of frequency can be approximated as

\begin{equation}
    h_0(f) = 2\frac{\sqrt{S_n(f)}}{(\Tfft \Tobs)^{1/4}}.\label{eqn:semi-coh-monochrom-sens}
\end{equation}
Here, $\Tfft=1800$ s in \cite{KAGRA:2021una} irrespective of the frequency, though this choice is primarily a function of computational cost, and $\Tobs$ is the observation time (in O3a, 6 months).

Plugging in $h_0(f)$ into Eq. \ref{eqn:h0(t)}, we can obtain the distance reach to \gwh systems slowly inspiraling towards one another

\begin{equation}
    d(f) =2\frac{(\Tfft \Tobs)^{1/4}}{\sqrt{S_n(f)}}\left(\frac{G \mathcal{M}}{c^2}\right)^{5/3}\left(\frac{\pi f_{\rm gw}(t)}{c}\right)^{2/3},
\end{equation}

From the distance reach, we can compute the space-time volume $\avgVT$, which, for nearby sources, is simply the volume of a sphere, whose radius is the distance reach, multiplied by the source duration in a particular frequency range $[f,f+\delta f]$:

\begin{equation}
    \avgVT = \frac{4}{3}\pi d(f)^3 T
\end{equation}
where $T=$max$(\Tobs,\Delta T)$, and \(\Delta T\) is the time spent by the binary system in a given frequency range $\delta f$: 
\begin{eqnarray}
  \Delta T &=& \frac{5}{256}\pi^{-8/3}\left(\frac{c^3}{G\mathcal{M}}\right)^{5/3} \left[f^{-8/3}-(f+\delta f)^{-8/3 }\right]\\ ~ 
  &\approx& \frac{5}{96}\pi^{-8/3}\left(\frac{c^3}{G\mathcal{M}}\right)^{5/3} \delta f f^{-11/3}
\label{eqn:deltaT}
\end{eqnarray}
where $\delta f=\dot{f}_{\rm max}\Tobs$. This means that

\begin{equation}
    \avgVT=\frac{5 \pi^{1/3} \dot{f} G^4 \mathcal{M}^4 \Tobs (\Tfft \Tobs)^{3/4} \left(\frac{c^3}{G \mathcal{M}}\right)^{2/3}}{9 c^9 f^{5/3} {S_n^{3/2}(f)}}
\end{equation}
Now, the number of binaries detectable at a given frequency is:

\begin{equation}
    N_{\rm bin}=\avgVT \mathcal{R},\label{eqn:nbin}
\end{equation}
where $\mathcal{R}$ is the formation rate density of binary \pbhs. Summing over all possible binaries detected at each frequency

\be
N_{\rm bin}^{\rm tot} = \sum_{ i} N_{\rm bin} (f_i)~.
\label{eqn:ntot}
\ee
and solving for $\mathcal{R}$, we arrive at

\be
\mathcal{R}=\frac{3}{4\pi}\left(\sum_i \avgVT(f_i)\right)^{-1}.
\label{eqn:ratedenssolved}
\ee
These rate densities are a function of each chirp mass to which we could be sensitive.

From the rate densities in Eq. \ref{eqn:ratedenssolved}, we can equate this quantity to analytical models for formation rate densities of \pbhs for equal-mass binaries:

\begin{eqnarray}
\mathcal{R} =& 1.04 \times 10^{-6}\, \mathrm{kpc}^{-3} \mathrm{yr}^{-1} f_{\rm sup} f(m_{\rm PBH})^2 
\left(\frac{m_\mathrm{PBH}}{M_\odot}\right)^{-32/37}  \left(f_{\rm PBH}\right)^{53/37},
\label{eqn:rate}
\end{eqnarray}
 and also asymmetric-mass ratio binaries, with $m_2\ll m_1$:

 \begin{eqnarray}
\mathcal{R} =& 5.28 \times 10^{-7}\, \mathrm{kpc}^{-3} \mathrm{yr}^{-1} f_{\rm sup} f(m_1) f(m_2) 
&\left(\frac{m_1}{M_\odot}\right)^{-32/37} \left(\frac{m_2}{m_1}\right)^{-34/37} \left(f_{\rm PBH}\right)^{53/37}~.
\label{eqn:rate_asymm}
\end{eqnarray}

We can constrain asymmetric mass-ratio binaries because the \cwh search is primarily sensitive to the chirp mass of the binary, so $m_1$ and $m_2$ can be chosen freely as long as $\mathcal{M}$ remains the same (assuming negligible eccentricity). Additionally, asymmetric mass-ratio binaries are more likely to occur. 

\begin{figure}[t]
\includegraphics[width=0.47\columnwidth]{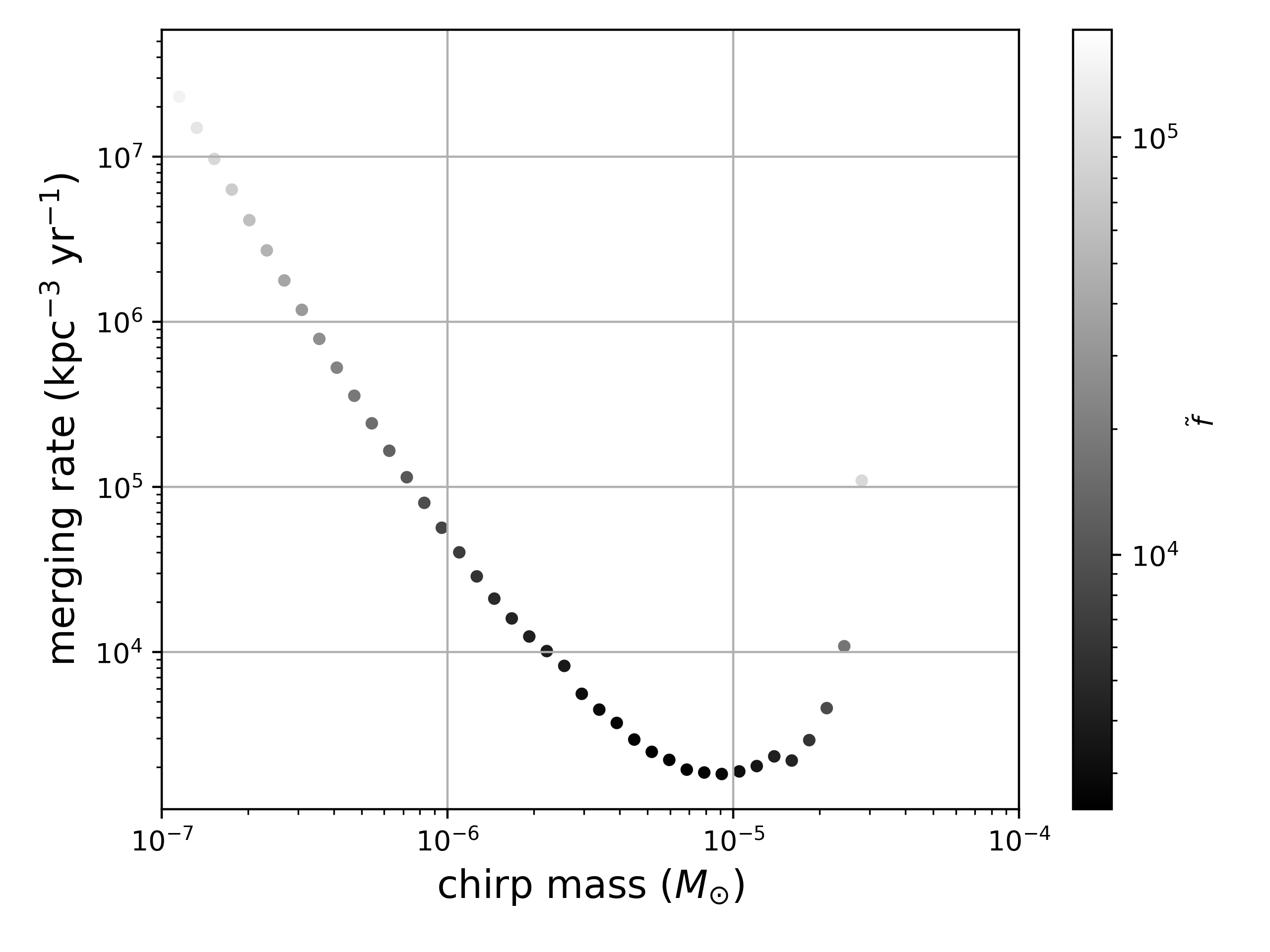}
\quad
\includegraphics[width=0.47\columnwidth]{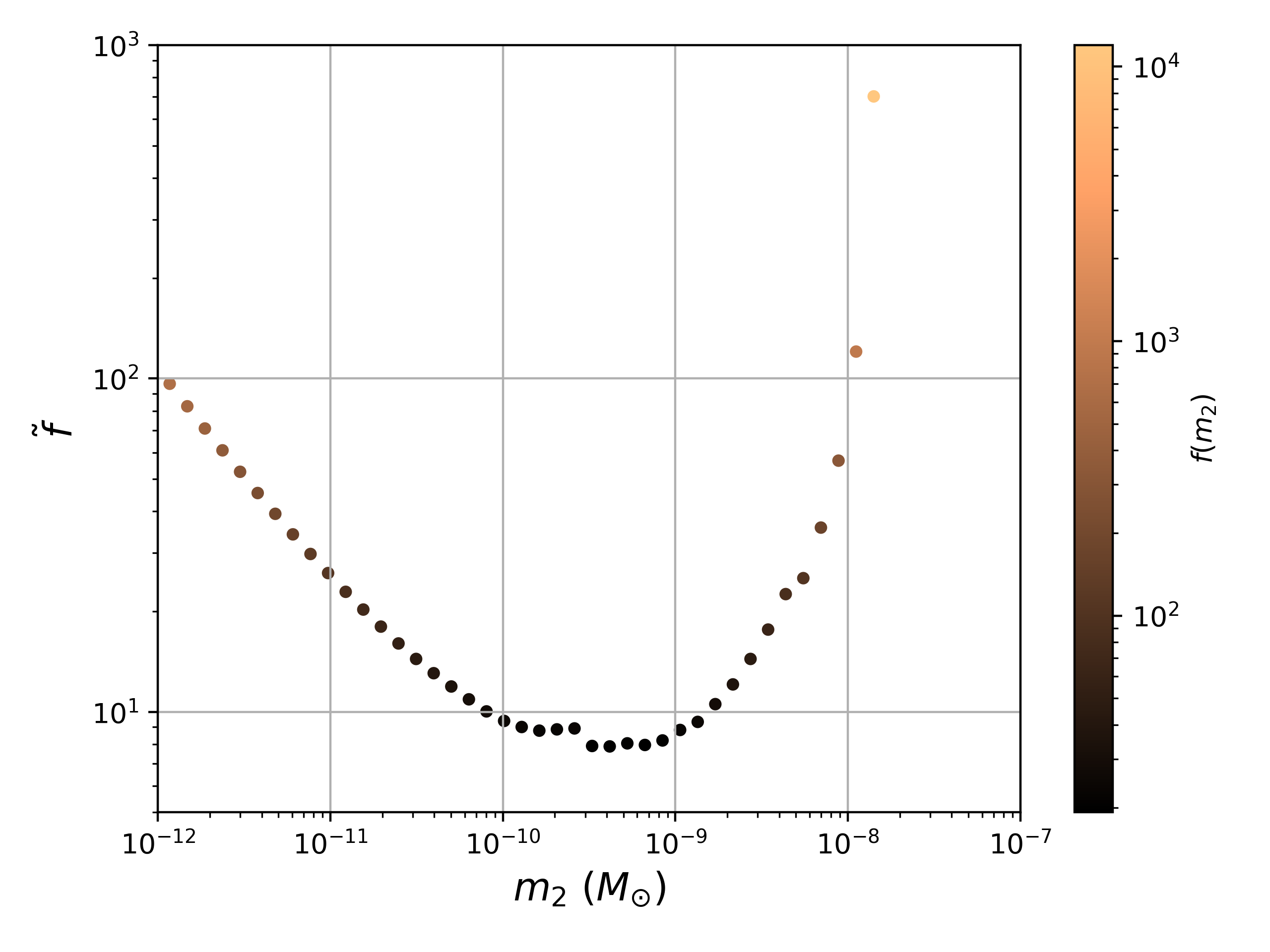}
\caption{Left: Upper limit on merging rates of planetary and asteroid chirp mass binaries. These rates do not depend on any particular PBH formation model; they only depend on the sensitivity of the \cwh search, i.e. the distance reach. Also shown in color are upper limits on the fraction of dark matter that primordial black holes could compose assuming rate models for equal-mass primordial black hole binaries. Right: Upper limit on the PBH abundance for asymmetric mass ratio binaries with one component at \(m_1 = 2.5 M_\odot\). Also shown in color is upper limit on the mass function of the smaller companion PBH assuming \(f(m_1) = 1\) and \(f_{\rm PBH} =1\). Taken from \cite{Miller:2021knj}}
\label{fig:rates}
\end{figure}

Since \pbhs are well-motivated by observations of merging black holes in the stellar-mass range, we consider systems with $m_1 = 2.5\, M_\odot$, as motivated by the QCD transition~\cite{Byrnes:2018clq,Carr:2019kxo,Clesse:2020ghq} and some mergers such as GW190425 and GW190814~\cite{Clesse:2020ghq}, in a binary with a much lighter \pbh of mass $m_2$. 
Combining Eq. \ref{eqn:ratedenssolved} with Eq. \ref{eqn:rate} and Eq. \ref{eqn:rate_asymm}, and denoting a model-independent parameter $\tilde f^{53/37} \equiv f_{\rm sup} f(m_1) f(m_2) f_{\rm PBH}^{53/37}$, we can arrive at constraints present in Fig. \ref{fig:rates}. These constraints, while not yet probing a physical regime, indicate that, soon, \cwh could detect nearby \pbh systems.

\subsection{Projected constraints for planetary-mass \pbhs}

Analogously to what was described in Sec. \ref{sec:asteroid-mass-constraints}, we can also constrain higher-mass with the semi-coherent methods described in \ref{sec:meths}. For systems with chirp masses $\mathcal{M}\in [10^{-5},10^{-2}]$, the linear approximation for the frequency/time evolution used in Sec. \ref{sec:asteroid-mass-constraints} breaks down; thus, we need to use the Generalized frqeuency-Hough transform or another semi-coherent method that can track the frequency evolution given in Eq. \ref{eqn:f-of-t}.

\begin{figure}[t]
    \centering
    \hspace{-3mm} 
    \includegraphics[width=\columnwidth]{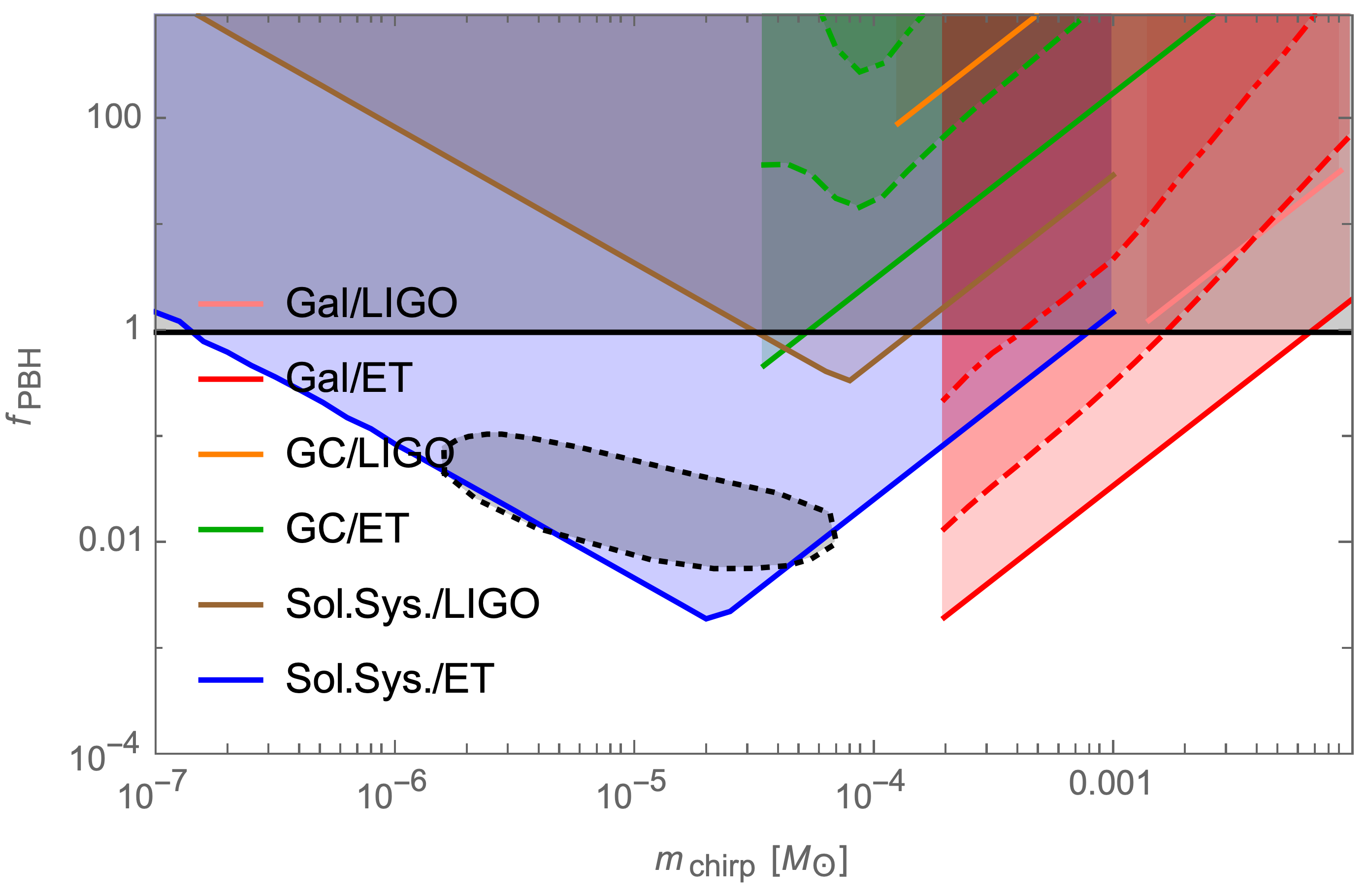}
    \caption{Expected limits on the dark matter fraction made of PBHs as a function of the chirp mass, for primordial binaries in Case 1 - agnostic mass function - (solid lines) and Case 2 - thermal mass function - (dashed lines), and for binaries formed through tidal capture in Case 2 (dotted-dashed lines).  The different colors represent the limits from galactic binaries (gal), from the galactic center (GC) and in the solar system vicinity (sol. sys.), for the expected sensitivities of advanced LIGO and ET. The dotted elliptic region represents constraints from the Optical Gravitational Lensing Experiment (OGLE) and the Subaru Hyper Suprime-Cam (HSC) \cite{Niikura:2019kqi} for comparison. Taken from \cite{Miller:2020kmv}. }\label{fig:limits}
\end{figure}

The procedure to obtain (projected) constraints for systems in this mass regime is almost identical to that in \ref{sec:asteroid-mass-constraints}, except that Eq. \ref{eqn:semi-coh-monochrom-sens} is replaced by one that accounts for the power-law evolution of the signal that also depends on the method used, and in Eq. \ref{eqn:deltaT}, the approximation that $\delta f/f\ll 1$ cannot be made. This equation can be found in \cite{Miller:2020vsl}, Eq. 29--31. Then, noting that $\Tfft$ is much smaller than in the previous case (because $\dot{f}$ is much bigger), the same procedure can be applied. We show the projected constraints on $f_{\rm PBH}$ for equal-mass systems in Fig. \ref{fig:limits}.

\subsection{mini-EMRIs}
A class of asymmetric-mass ratio binaries could be so-called mini extreme-mass ratio inspirals (mini EMRIs), which could be detected by current \gwh interferometers \cite{Guo:2022sdd}. These systems could be composed of a $10\msun$ or $100\msun$ primary object whose companion would be lighter by a factor of $\sim 10^5$, i.e. $m_2/m_1=q=10^{-5}$. Such systems would require the lighter object to be a \pbh or another exotic compact object.  

\begin{figure*}[t]
\includegraphics[width=0.49\textwidth]{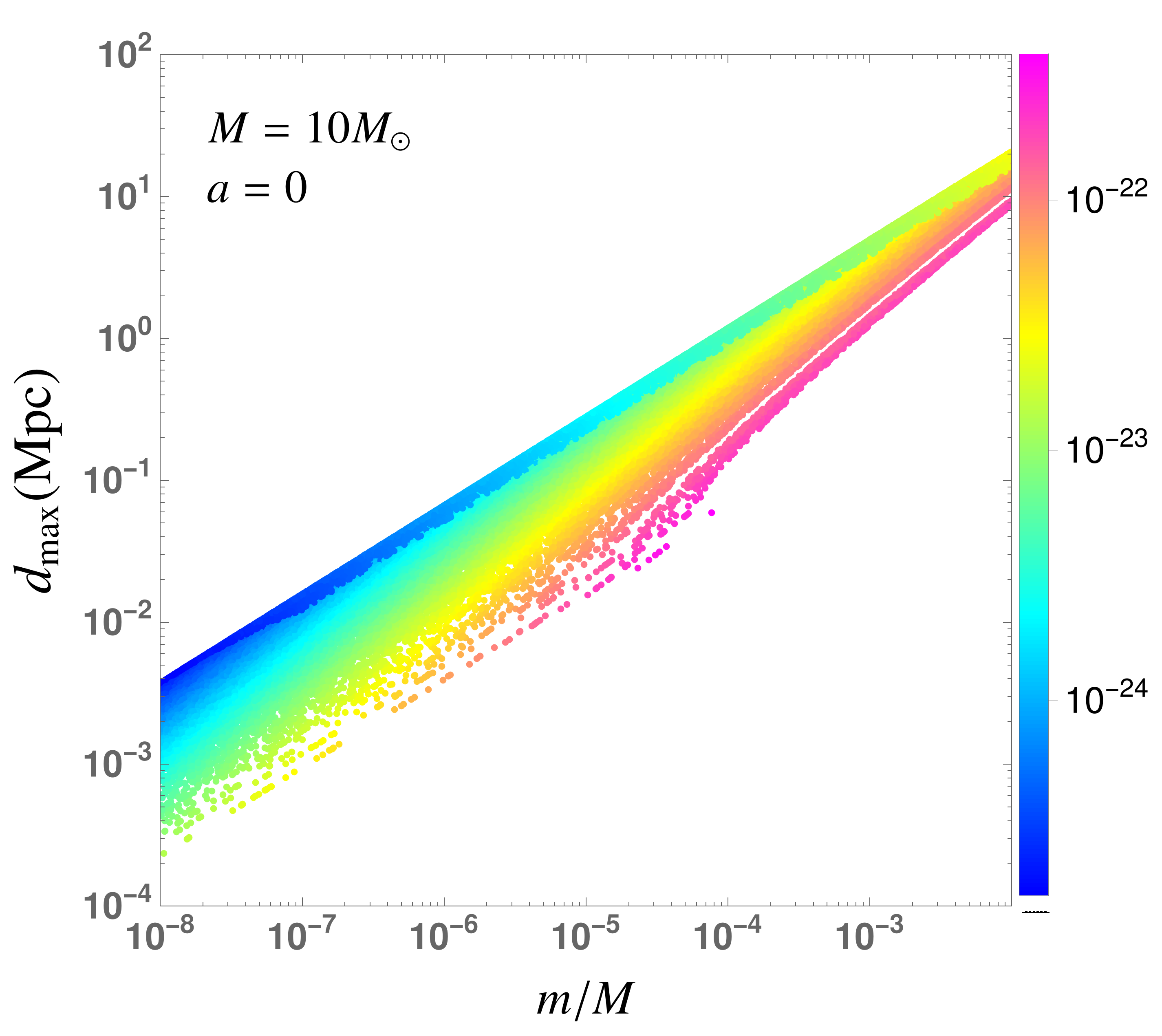}
\quad
  \includegraphics[width=0.435\textwidth]{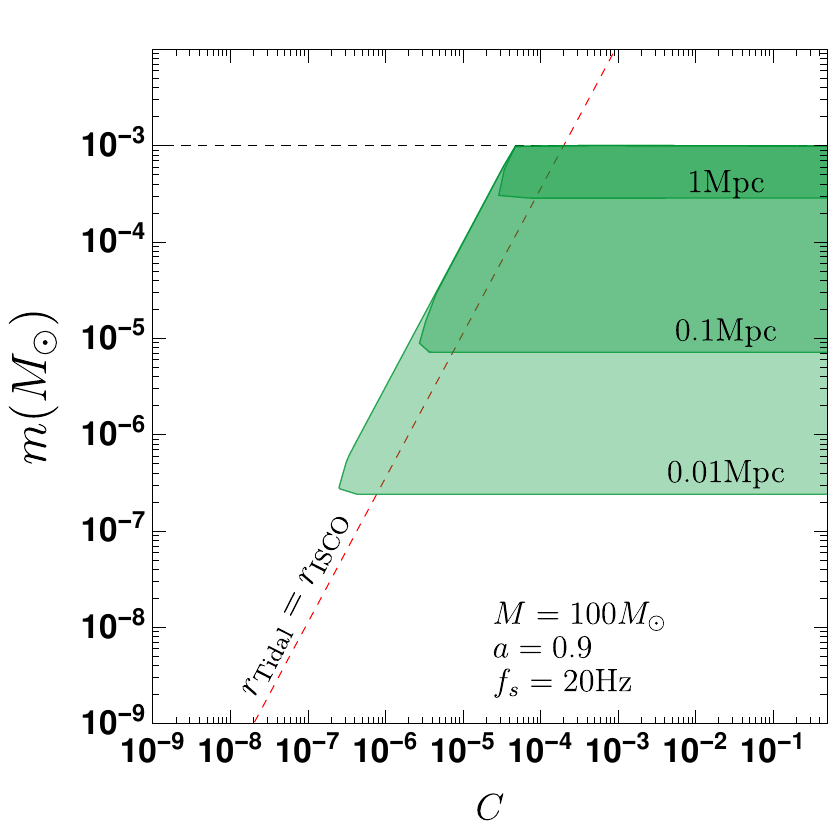}
  \caption{\label{fig:res}
  Left: estimated distance reach at 95\% confidence as a function of mass ratio, with the minimum detectable \gwh amplitude colored. Right: Sensitivity to a mini-EMRI consisting of a $100 M_{\odot}$ central massive compact object and an ECO specified by its mass (vertical axis) and compactness (horizontal axis), where
  the color-shaded regions denote where the corresponding mini-EMRI system can emit \gws and be detected by LIGO, assuming $CR_{\rm thr}=5$. Taken from \cite{Guo:2022sdd}.}
\end{figure*}

Though similar to the previous discussions of asymmetric mass-ratio \pbhs, the signal model is slightly different:

\begin{align}
\frac{df}{dt} = k f^{11/3} C_f(a, f) ,\label{eq:fdot-mini-emri}
\end{align}

\begin{eqnarray}
h_0 &=& \frac{4}{d} \left(\frac{G M_c}{c^2}\right)^{5/3} \left(\frac{\pi f}{c}\right)^{2/3} C_h (a, f) .
\end{eqnarray}
where these two factors, $C_f(a,f)$ and $C_h(a,f)$ characterize the change in the orbital (or \gwh) frequency accounting for relativistic effects. Plots of these two quantities are shown in \cite{Guo:2022sdd}. The luminosity distance reach for these kinds of systems varies depending on the mass ratio, and is shown in Fig. \ref{fig:res}, along with projected constraints on the exotic companion as a function of its compactness

Though the relativistic factors imply a variation in the frequency evolution over time in Eq. \ref{eqn:f-of-t}, if we wish to apply the Hough Transform to look for these systems, we can treat Eq. \ref{eq:fdot-mini-emri} as an ``effective'' power-law, so that its form resembles in Eq. \ref{eq:fdot-mini-emri} that, though not exactly $f^{11/3}$. In a real search, therefore, this would require searching over a range of power laws, in addition to a range of chirp masses and frequencies. Still, however, the system could not be too close to the plunge in order to employ this approach. A better idea may be the Viterbi method that is not bound to a particular model and could potentially pick out mini \emri signals where $C_f(a,f)\gg 1$.

\section{EMRIs}

\subsection{Motivation}

Extreme mass ratio inspirals (EMRIs) consist of a primary mass $m_1\sim\mathcal{O}(10^9)\msun$ (such as a supermassive \bh at the center of our galaxy) and a secondary mass $m_2\sim \mathcal{O}(1)\msun$ (a stellar-mass compact object such as a neutron star, black hole or white dwarf). Ordinary stellar objects would be tidally disrupted by such a supermassive \bh; thus, EMRIs require that the secondary object is compact \cite{Maggiore:2018sht}. In contrast to the binary \bhs that merge in less than a second in ground-based observatories, \emris would be visible in space-based \gwh detectors at mHz frequencies, and spend years in band before the secondary object plunges into the supermassive \bh. As the secondary objects inspirals into the primary one, a significant amount of \gwh power will be emitted at low frequencies.

Such systems are interesting from both astrophysics and fundamental physics perspectives \cite{Babak:2017tow}. The detection of such sources would allow us to test whether the supermassive \bhs indeed behave as Kerr \bhs \cite{Barack:2006pq,Gair:2012nm}, measure the mass distribution of supermassive \bhs and their stellar environments \cite{Gair:2010yu}, measure the background spacetimes of supermassive \bhs \cite{Glampedakis:2005cf}, and measure the presence of gas around the primary object \cite{Amaro-Seoane:2007osp}.

To obtain a sense of how long these signals will last, we can integrate Eq. \ref{eqn:f-of-t} to obtain the number of cycles 

\begin{eqnarray}
    N_{\rm cyc} &=& \int_{t_0}^{t} f(t)dt \\
    &=& \int_{\fmin}^{\fmax} \frac{f}{\dot{f}} df
\end{eqnarray}
and also add a redshift-dependence $1+z$ for binaries at cosmological distances \cite{Maggiore:2018sht}

\begin{eqnarray}
    N_{\rm cyc} &=& \frac{1}{32\pi^{8/3}} \l\frac{G\mathcal{M}}{c^3}\r^{-5/3}\l \fmin^{-5/3}-\fmax^{-5/3}\r \\
    &=&10^5 \l \frac{2\text{ mHz}}{\fmin}\r^{5/3} \l\frac{10^3\msun}{\mathcal{M}}\r^{5/3}\l\frac{2}{1+z}\r^{5/3}, \nonumber
\end{eqnarray}
where $\mathcal{M}=10^3\msun$ could correspond to $m_1=10^6\msun$ and $m_2=10\msun$.

To compare $N_{\rm cyc}$ to that of systems visible in ground-based detectors, we see that we can set $m_1=m_2=10\msun$ ($\mathcal{M}\simeq 8.7\msun$) at a frequency of 10 Hz, we obtain $N_{\rm cyc}\simeq 600$ at a redshift of 1. An \emri, therefore, spends orders of magnitude longer in the frequency band of space-based detectors compared to its duration in ground-based ones.

\subsection{Search techniques}

As discussed in previous sections, matched filtering provides the best sensitivity of any data analysis method. However, creating \emri waveforms may pose problems in the future. First, the number of cycles over which the template has to be reasonably accurate is much higher than in ground-based detectors. Second, because the secondary object will reach speeds close to the speed of light, Post-Newtonian expansions cannot be employed to derive the waveform of these systems, nor is numerical relativity feasible, since it only works for mass ratios up to $10^5$ and $\sim 10^2$ cycles. It is possible, therefore, that semi-coherent methods discussed in Sec. \ref{sec:meths}, which operate by breaking data into smaller coherent segments that are analyzed incoherently in the time/frequency plane, may be used in the future to search for \emris. Furthermore, if we do wish to employ matched filtering, some approaches have already been designed, based on self-force \cite{Maggiore:2018sht}, that are able to expand the waveform in powers of the (small) mass ratio $m_2/m_1$ of \emris. While all these techniques seem promising, they are in their infancy and need to be extensively studied to handle the \emri case.

\subsection{Detection prospects}

The number of \emris as a function of redshift depends on a variety of astrophysical unknowns: (1) the number of supermassive \bhs formed as a function of their mass, spin and redshift, (2) the properties of the stars around the supermassive \bh: whether the supermassive \bh is surrounded by stars spatially distributed along a cusp around it (necessary for \emri formation), how galaxy mergers over cosmic time have affected the cusp (mergers would erode the cusps on some timescale, and on a different timescale, the cusp could regrow, (3) how many plunges versus \emris there are, (4) the duty cycle of \emris, and (5) the chosen $m_2$ \cite{Babak:2017tow}. When models for each of these are chosen, the rate of \emri formation can be derived, which can range from $\mathcal{O}(10)$ to $\mathcal{O}(10^5)$ per year up to a redshift of $z=4.5$, of which only 10\% to 50\% of them would be detectable by LISA, depending on the waveform -- see Tab. 1 of \cite{Babak:2017tow}.

\begin{figure*}[htb]
\centering
    \includegraphics[width=0.47\columnwidth]{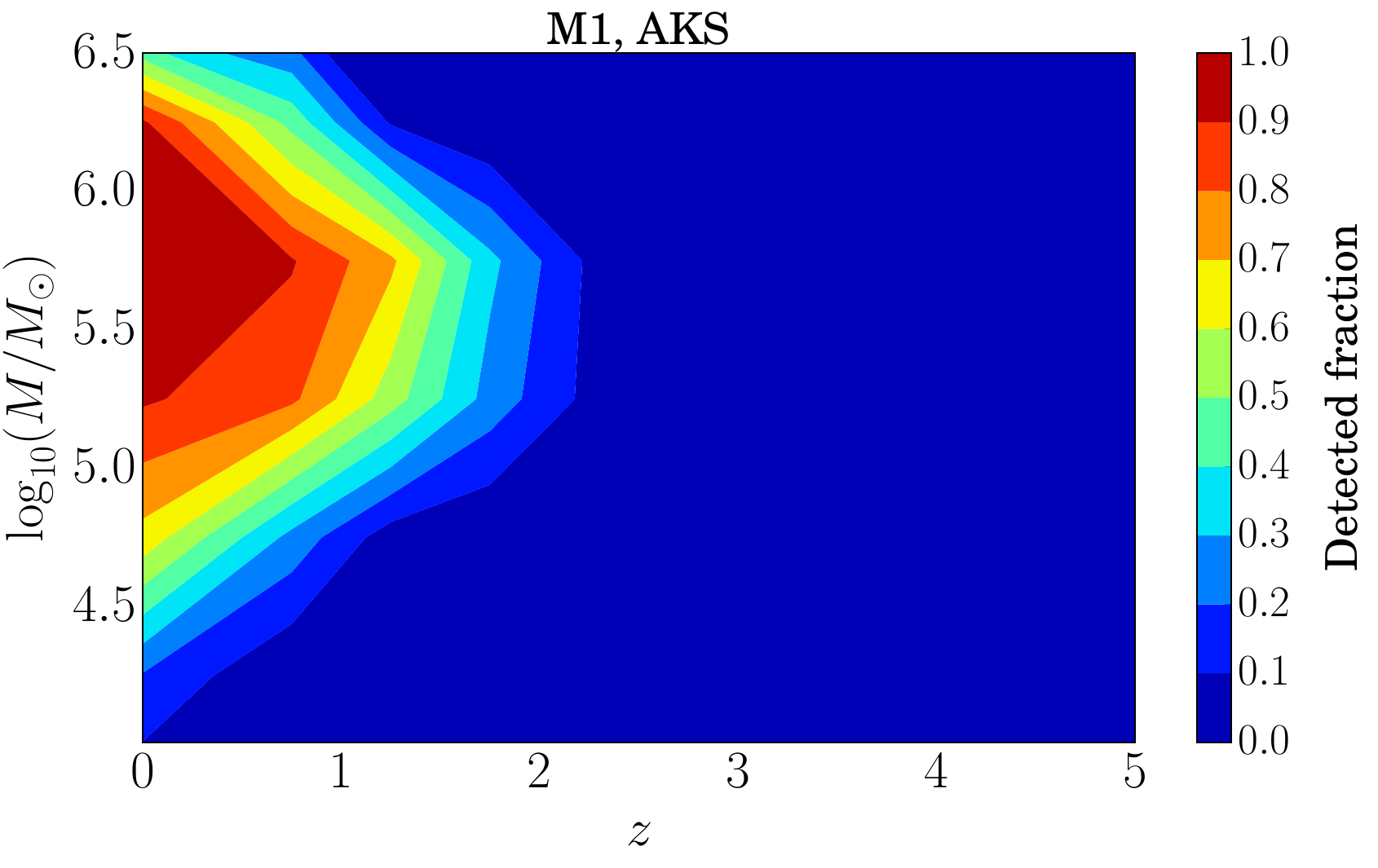} \quad \includegraphics[width=0.47\columnwidth]{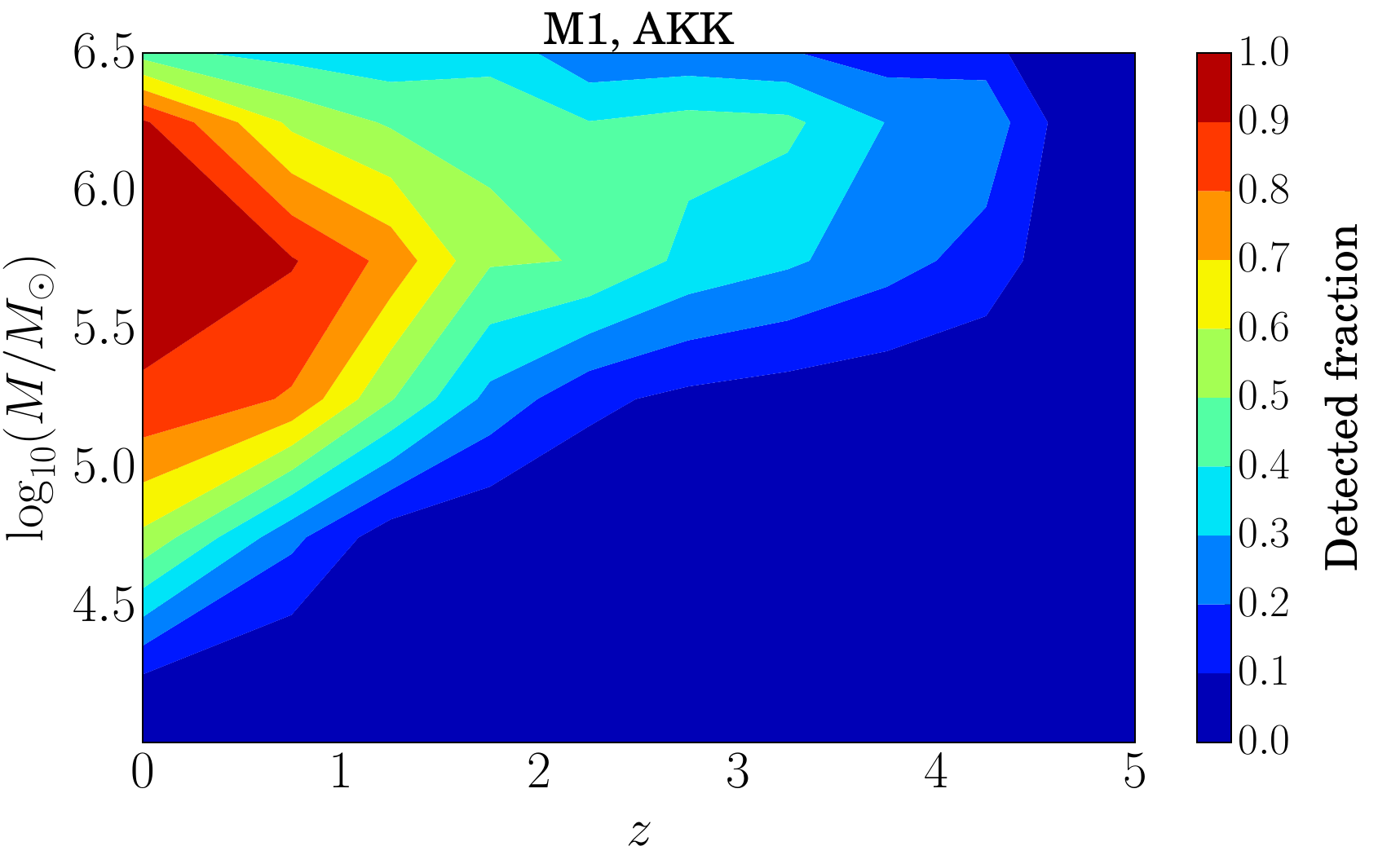}  \\
     \includegraphics[width=0.47\columnwidth]{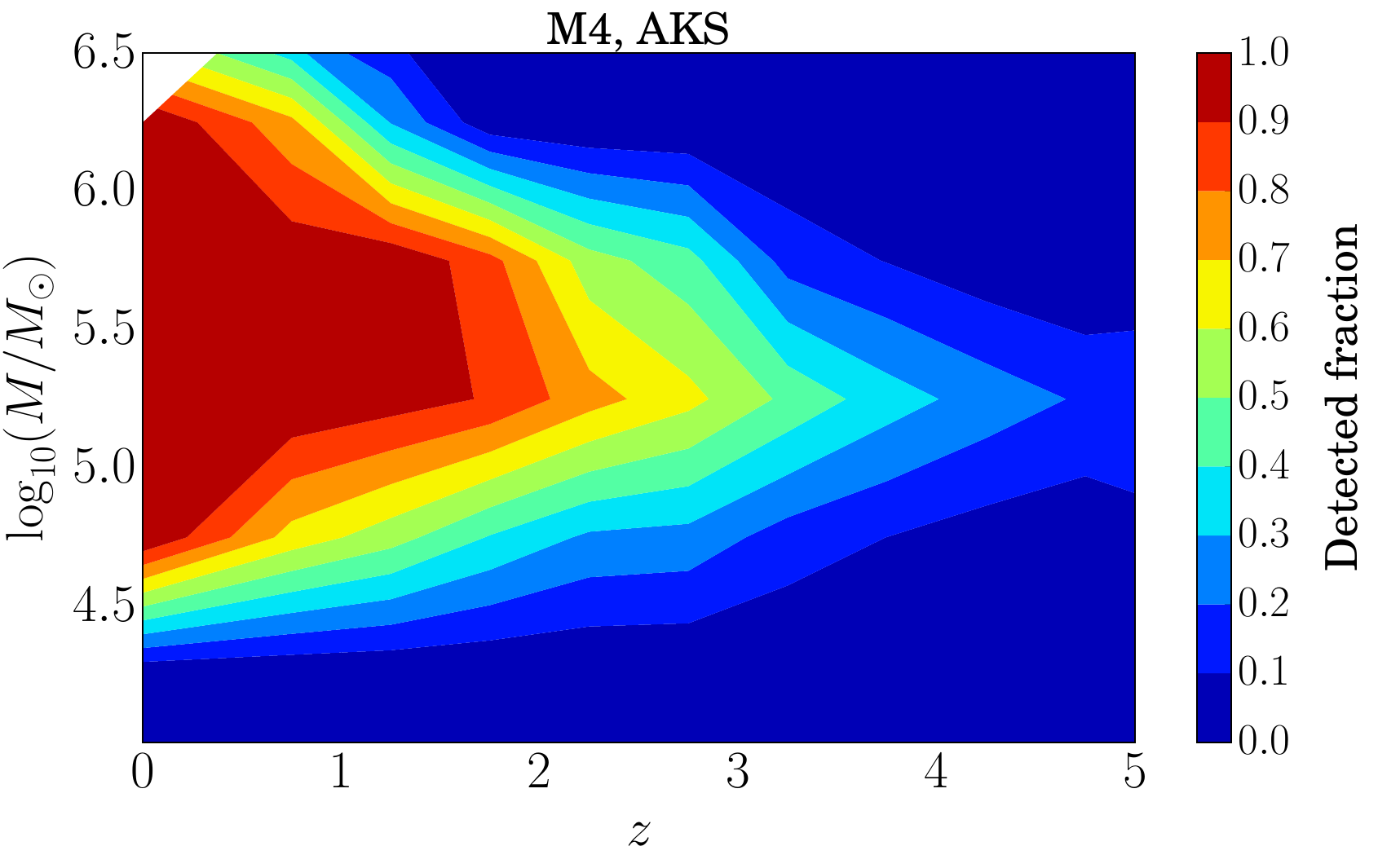} \quad \includegraphics[width=0.47\columnwidth]{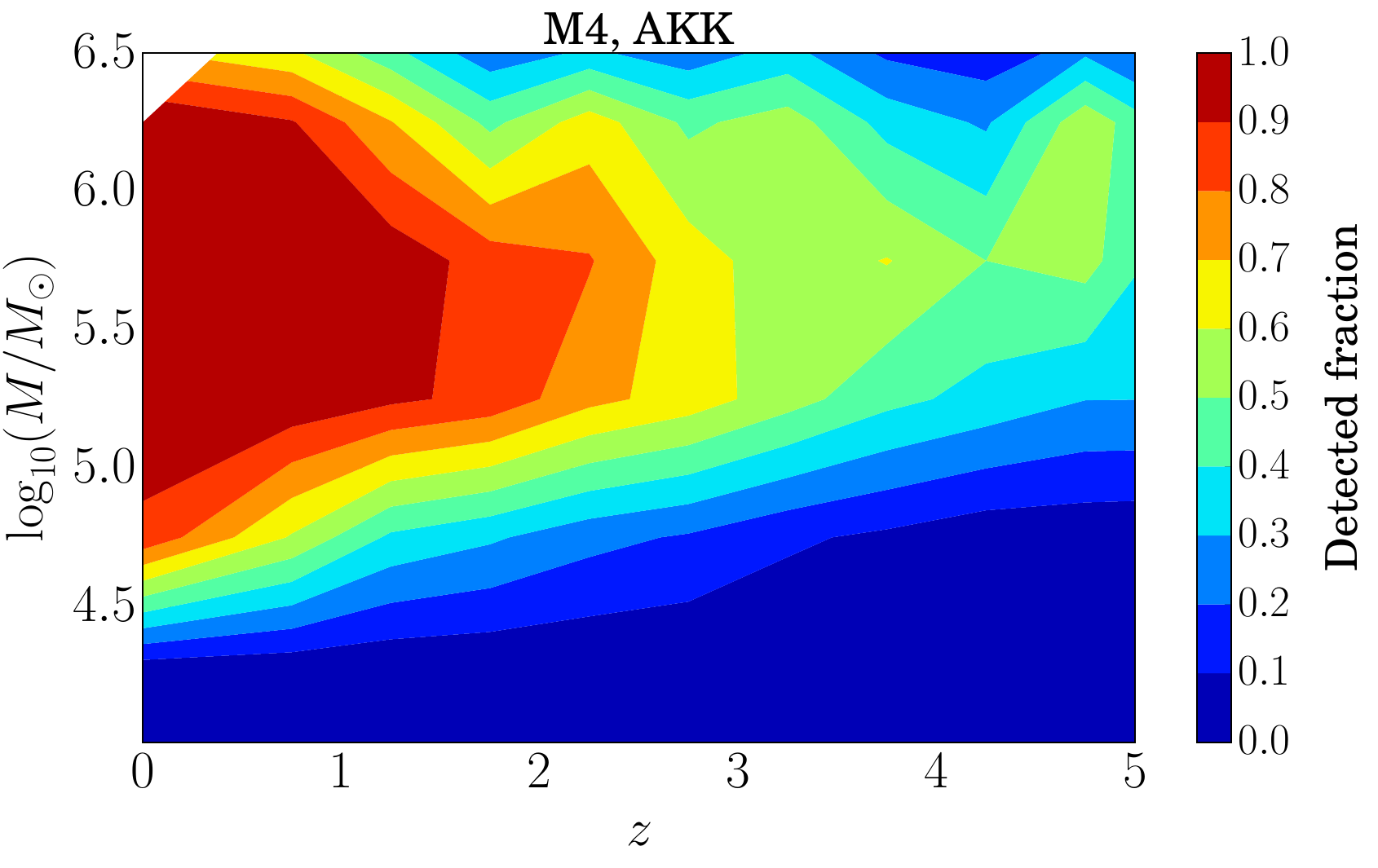}  \\ 
  \caption{Fraction of the intrinsic EMRI population detectable by LISA as a function of source-frame total mass and redshift, for two models of supermassive \bh mass function, spin, cusp regrowth time, etc. (M1 and M4) and with two waveforms constructed to perform a matched filtering analysis (AKS and AKK). Taken from \cite{Babak:2017tow}.}
\label{fig:det_frac}
\end{figure*}

In Fig. \ref{fig:det_frac}, we show the fraction of detectable \emri events using two waveforms in a matched filtering analysis at LISA sensitivity as a function of primary component mass and redshift, fixing $m_2=10\msun$ (M1) and $m_2=30\msun$ (M4). We can see that locally, i.e. $z<1$, we would be able to detect close to all the \emris around, and we observe a fall-off with redshift.

\section{Conclusions}

We have described in great detail how searches for \ssm \pbhs from predominately current-generation ground-based work. Such a mass range also presents exciting opportunities for future observatories on the ground as well, such as Einstein Telescope and Cosmic Explorer \cite{Punturo:2010zz,Reitze:2019iox}, and space-based ones, such as DECIGO, LISA, Taiji and TianQin \cite{Kawamura:2006up,LISA:2017pwj,Ruan:2018tsw,Luo:2019zal}. Indeed, in future ground-based detectors, the lowest detectable frequency of such inspiraling systems will be much around $2-5$ Hz instead of the $10-15$ Hz it currently is, which will allow a significantly longer observation time, and potentially the localization of such systems using only the relative motion of the earth and source. Furthermore, with roughly a factor of $\sim 10$ improvement in the luminosity distance reached, we will be able to see \pbhs inspiraling at least in a 1000x larger volume than we currently do. The expected exquisite sensitivity of these detectors, as well as long-duration sources, will require surmounting new challenges related to computational cost, estimation of the signal-free background, signals occurring simultaneously, and non-stationary noise. The methods presented here, especially those from tCW and CW searches, are apt at handling these problems \cite{Miller:2023rnn}, and along with matched filtering, could constitute a major probe of sub-solar mass \pbhs in the future.

Regarding space-based detectors, such systems could easily spend years visible mHz or deci-Hz frequencies, enabling the possibility of performing multi-band \gwh astronomy \cite{Sesana:2016ljz}. In essence, the source parameters could be estimated using space-based detector data, and further measurements of the inspiral could be made in future ground-based detectors. Potentially, if ultra-high frequency \gwh detectors come online, the mergers of such \ssm \pbhs could be detected as well \cite{Aggarwal:2020olq}.

The future is therefore bright for searching for and hopefully finding \gws from binary \ssm \pbhs.

%
\addcontentsline{toc}{section}{Appendix}
 \bibliographystyle{unsrt}
 \bibliography{authorsample.bib}

\end{document}